\documentclass[twocolumn]{aastex6}

\usepackage{float}
\usepackage{amsmath}
\usepackage{color}

\newcommand{\cn}[1]{{\color{blue} [refs]}}

\newcommand{\fc}[1]{{\emph{(add cite to bibliography)}}}
\newcommand{\KpIp}[1]{{Kirkpatrick (2015)}}
\newcommand{\CHIp}[1]{{Hayward et al. (\emph{in prep})}}
\newcommand{\ERIp}[1]{{Roebuck et al. (\emph{in prep})}}
\newcommand{\nd}[1]{{\emph{***Not done...}}}
\newcommand{\agnf}[1]{$L_{\rm{AGN}} / L_{\rm{IR}}$}
\newcommand{\lir}[1]{$L_{\rm{IR}}$}
\newcommand{\emit}[1]{\rm{emit}}
\newcommand{\obs}[1]{\rm{obs}}
\newcommand{\data}[1]{\rm{data}}
\newcommand{\model}[1]{\rm{model}}
\newcommand{\um}{\,$\mu$m}
\newcommand{\lsun}{\,$\rm{L}_{\odot}$}
\newcommand{\rcom}{}
\newcommand{\rcomtwo}{}

\newcommand{\lge}{M2}
\newcommand{\vc}{M4}
\newcommand{\ce}{M6}

\newcommand{\mir}{$f\rm{(AGN)_{MIR}}$}

\newcommand{\tir}{$f\rm{(AGN)_{IR}}$}

\newcommand{\fagnbol}{f\rm{(AGN)_{bol}}}

\newcommand{\inp}{$f\rm{(AGN)_{bol}}$}

\newcommand{\hosteqn}{f\rm{(AGN)_{IR}}}
\newcommand{\host}{$f\rm{(AGN)_{IR}}$}
\newcommand{\iremp}{$f\rm{(AGN)_{IR,emp}}$}

\begin{document}
\title{The Role of Star-Formation and AGN in Dust Heating of z=0.3-2.8 Galaxies - II. Informing IR AGN fraction estimates through simulations}
\author{Eric Roebuck\altaffilmark{1}, Anna Sajina\altaffilmark{1}, Christopher C. Hayward\altaffilmark{2,3}, Alexandra Pope\altaffilmark{4}, Allison Kirkpatrick\altaffilmark{4,5}, Lars Hernquist\altaffilmark{3}, Lin Yan\altaffilmark{6}}
\altaffiltext{1}{Department of Physics and Astronomy, Tufts University, Medford, MA 02155, USA; eric.roebuck@tufts.edu} 
\altaffiltext{2}{Center for Computational Astrophysics, 160 Fifth Avenue, New York, NY 10010, USA} 
\altaffiltext{3}{Harvard-Smithsonian Center for Astrophysics, 60 Garden Street, Cambridge, MA 02138, USA}
\altaffiltext{4}{University of Massachusetts Amherst, Amherst, MA 01003, USA} 
\altaffiltext{5}{Yale Center for Astronomy and Astrophysics, Physics Department, P.O. Box 208120, New Haven, CT 06520, USA} 
\altaffiltext{6}{Infrared Processing and Analysis Center, California Institute of Technology, Pasadena, CA 91125, USA}

\begin{abstract}
A key question in extragalactic studies is the determination of the relative roles of stars and AGN in powering dusty galaxies at $z$\,$\sim$\,1-3 where the bulk of star-formation and AGN activity took place. In Paper I, we present a sample of $336$ 24\,$\mu$m-selected (Ultra)Luminous Infrared Galaxies, (U)LIRGs, at $z \sim 0.3$-$2.8$, where we focus on determining the AGN contribution to the IR luminosity. Here, we use hydrodynamic simulations with dust radiative transfer of isolated and merging galaxies, to investigate how well the simulations reproduce our empirical IR AGN fraction estimates and determine how IR AGN fractions relate to the UV-mm AGN fraction.
We find that: 1) IR AGN fraction estimates based on simulations are in qualitative agreement with the empirical values when host reprocessing of the AGN light is considered;
2) for star-forming galaxy-AGN composites our empirical methods may be underestimating the role of AGN, as our simulations imply $>$\,50\,\% AGN fractions, $\sim$\,3$\,\times$ higher than previous estimates;
3) 6\% of our empirically classified ``SFG" have AGN fractions $\gtrsim$ 50\%. While this is a small percentage of SFGs, if confirmed, would imply the true number density of AGN may be underestimated;
4) this comparison {\rcomtwo depends} on the adopted AGN template -- those that neglect the contribution of warm dust lower the empirical fractions by up to 2$\times$; and
5) the IR AGN fraction is only a good proxy for the intrinsic UV-mm AGN fraction when the extinction is high ($A_V\gtrsim 1$ or up to and including coalescence in a merger).
\end{abstract}

\section{Introduction}

Understanding galaxies at $z \sim 1$-$3$ is of key importance to galaxy evolution studies because both the star formation rate density \citep[see][for a recent review]{MadauDickinson2014} and quasar number density \citep{Richards2006b} peak at this epoch. Along with the $M_{\rm{BH}}-\sigma$ relation \citep{Ferrarese2000}, these observations suggest that the accumulation of stellar mass and growth of super-massive black holes are closely tied \citep[see e.g.][]{Hopkins2006, Hopkins2008}. The increase in number density of luminous and ultraluminous infrared galaxies (LIRGs and ULIRGs respectively) up to $z \sim 2$ makes them the dominant contributor to the SFR density peak \citep{Murphy2011,Magnelli2011,Casey2012}. However, understanding exactly how much star formation takes place in such systems requires accurate determinations of the fraction of their power output that is due to recent star formation rather than AGN. The high levels of obscuration in such galaxies make answering this question notoriously difficult. 
Analysis of the infrared spectral energy distribution (IR SED), especially mid-IR spectra when available, has been our best tool to determine the level of AGN activity in such heavily dust obscured systems \citep[e.g.][]{Armus2007,Sajina2007,Yan2007,Pope2008,Veilleux2009,Kirkpatrick2012,Sajina2012}. 
The contribution of AGN to the IR luminosity\footnote{Throughout this paper, IR refers to the integrated 8-1000\um\ emission.} is typically referred to as the IR ``AGN fraction" or $f{\rm{(AGN)_{IR}}}$. Traditionally, determining $f{\rm{(AGN)_{IR}}}$ is based on assuming that the hot dust giving rise to the mid-IR continuum is exclusively due to an AGN torus, while the far-IR cold dust emission peak is entirely powered by stars \citep[e.g][]{Polletta2007,Sajina2007}. The warm dust ($\sim$\,80-100K) giving rise to the 20-40\um\ continuum is more uncertain as it can be due to star formation \citep[e.g.][]{Veilleux2009} {\rcom or to reprocessing in an NLR} \citep[e.g. see][for a review]{Netzer2015}. This can account for the typically greater warm dust component in empirical AGN templates \citep[e.g.][]{Richards2006a,Mullaney2011} relative to pure AGN torus models \citep[e.g.][]{Nenkova2008,Honig2010}. 

Aside from the uncertainty regarding the role of the NLR, this view ignores the fact that the AGN is embedded in its host galaxy, and the light from it is subject to further processing therein. The effects of this galaxy-scale dust processing of the AGN emission can be investigated by performing radiative transfer on hydrodynamical simulations of galaxies. 

The goal of this paper is to inform empirical IR AGN fraction estimates by comparing simulated and observed IR SEDs of a sample covering the redshift and luminosity regime most critical to the build-up of stars and black holes in the Universe. This paper is second in a series. Paper I \citep{Kirkpatrick2015} presents our sample of 343 24$\mu$m-selected $z\sim$0.3-2.8 (U)LIRGs with exceptional coverage from the optical to the far-IR/mm including \emph{Spitzer}/IRS mid-IR spectra. That paper includes a state-of-the-art spectro-photometric analysis of the observed IR SEDs yielding empirical \tir{}. In this paper (Paper II), we test whether simulated galaxies can reproduce the observed SEDs of our sample, which covers a wide range in IR AGN fractions; compare the empirical and simulation-based \tir{}; and investigate how such IR AGN fractions constrain the intrinsic AGN contribution to the power output of dusty galaxies. In Paper III (Roebuck et al., in prep.), we will present a more detailed comparison between the simulated and observed SEDs, including a discussion of the merger stage/morphology, gas fractions, star formation rates, and stellar masses of the galaxies.  

The structure of the paper is as follows. In Section~\ref{sec:sample} we describe the observed data. In Section~\ref{sec:simdata} we summarize the methodology underlying the \textsc{gadget}+\textsc{sunrise} simulations and present the details of our specific simulation library. In Section~\ref{sec:analysis} we use our suite of simulations to explore the dependence of IR AGN fraction estimates on the intrinsic AGN fraction, and on parameters such as merger stage, level of obscuration, initial gas fraction and viewing perspective. We then present a direct comparison between empirical SED-fitting based AGN fractions to those implied by the best fitting simulated SED to the observed SED. We include estimates of the systematic uncertainties in the derived AGN fractions. In Section~\ref{sec:discussion}, we discuss the caveats associated both with the simulation-based and empirical AGN fractions. We present the summary and conclusions in Section~\ref{sec:conclusions}. In the appendix, we investigate the potential dependence of our results on the assumed AGN SED template and the sub-resolution structure of the ISM of the simulated galaxies.

\begin{figure}[h]
   \centering
   \includegraphics[width=\columnwidth]{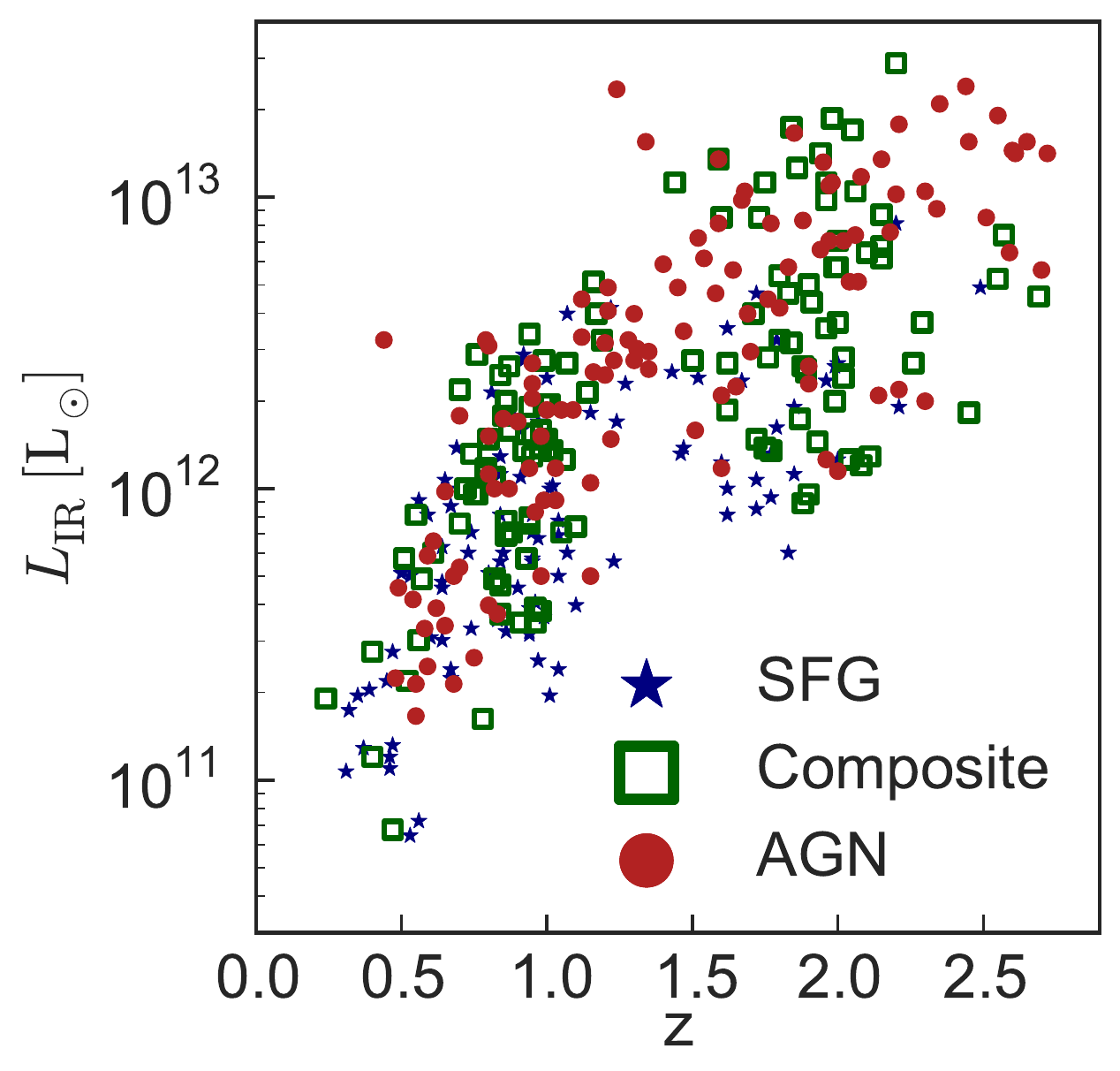}
   \caption{The luminosity-redshift distribution of our {\rcom observed} sample. The redshifts are based on optical, near-IR or mid-IR spectra. The luminosities are based on the IR SED fitting \citep{Kirkpatrick2015}. The three classes shown (SFG, Composite and AGN) are based on mid-IR spectral fitting as described in the text.}
   \label{fig:lumz}
\end{figure}

\section{Observed Data \& Empirical Classification} \label{sec:sample}

Figure~\ref{fig:lumz} shows the distribution of our sample of galaxies in redshift and IR luminosity. Full details on the sample selection and coverage are presented in \citet{Kirkpatrick2015}; here we provide a brief summary. The sample is representative of a 24\um\ flux limited-selection with $S_{24 \rm{\mu m}}$\,$>$\,0.1\,mJy. Most importantly, all galaxies in the sample have low resolution (R $= \lambda / \Delta \lambda \sim 100$) mid-infrared \emph{Spitzer} IRS spectroscopy\footnote{Refer to Paper I \citep{Kirkpatrick2015} for a complete list of the IRS programs involved.} \citep{houck2004}. In addition, our galaxies have up to 11 broadband photometric points across the IR SED that include \emph{Spitzer} IRAC in the near-IR, MIPS 24 and 70 $\ \rm{\mu m}$, and {\sl Herschel} PACS and SPIRE photometry \citep{Fazio2004,Rieke2004,Poglitsch2010,Griffin2010}. From the original sample of 343 galaxies in Paper I \citep{Kirkpatrick2015} we remove the six galaxies with z$>$2.8 as redshift determinations may be uncertain given our special coverage. We additionally remove one galaxy where the spectra may not match the photometry. Our final sample consists of 336 galaxies.

In Paper I, we fit the mid IR SEDs using a linear combination of: 1) a star forming component represented either as the local starburst component of \citet{Brandl2006} or the starburst M82 \citep{Forster2003}; and 2) an AGN component comprised of a power law, with the slope and normalization as free parameters. Each component has a separate screen extinction based on the MW extinction curve from \citet{Draine2003}. The mid-IR AGN fraction \mir{} is calculated by taking the ratio of the AGN component over the total SED integrated over $\sim$5-18$\ \rm{\mu m}$. 

We adopt the classification scheme from Paper I where star-forming galaxies (SFGs) have \mir{}$<0.2$, AGN have \mir{}$>0.8$, and composites have intermediate values of \mir{}$=0.2-0.8$. {\rcom \citet{Kirkpatrick2015} find 70\% of their sources have $| f {\rm (AGN)}_{\rm MIR}- f \rm{(AGN)}_{\rm MIR,unextinct}|<0.1$, with an average value of $\sim 0.06$. For this reason in this paper we use the unextincted AGN contribution.}

In Paper I, we construct full IR SED templates, based on the above mid-IR classification, and in bins of IR luminosity and redshift. These full IR SED templates, are in turn decomposed with a linear combination of a star-forming component and AGN (represented by the respective templates from \citet{Kirkpatrick2012}) to find a total 8-1000\um\ IR AGN fraction \tir{} for each template. We find that \mir{} correlates quadratically with \tir{} for the template SEDs. This relation is then applied to each source in our sample to derive individual empirical \tir{}. These values are denoted as \iremp{} throughout the rest of this paper. 

{\rcom For a given galaxy the empirical \host{} is consistent whether we use true photometry or one based on the best fit simulated SEDs as described in Section~\ref{sec:agnfrac}. We test this for three random galaxies (a SFG, a Composite, and an AGN), by taking the best-fit simulated SED, generating synthetic photometry (including IRS spectra) from it and running this ``simulated" photometry through the full analysis of \citet{Kirkpatrick2015}. The resulting mid-IR classifications were unchanged, and the 8-1000\um\ AGN fractions were within 20\% of the values inferred from the real photometry (well within the systematic uncertainties we derive in Section\,\ref{sec:agnfrac})}. 

A key result of Paper I is that the warm dust component is consistent with being 
AGN powered -- this is seen in particular in that the temperature of the warm dust increases as the mid-IR AGN strength increases. This empirical result is not proof, but is consistent with this warm dust being associated with an NLR as discussed in the Introduction. An even broader result is that AGN dominate the counts above $S_{24 \rm{\mu m}}>0.5$ mJy, but even down to the lowest flux levels (0.1\,mJy), sources with significant AGN (AGN+Composite classification) account for 40-60\% of the counts. Roughly half of these faint AGN, are in the Composite population, that would likely be missed by traditional AGN surveys. 

\section{Simulated Data} \label{sec:simdata}

We use a suite of idealized simulations of isolated disk galaxies and galaxy mergers {\rcom to compare to our observations}. All of the hydrodynamical simulations were presented originally in previous works (see Table\,\ref{tab:modeltable}), but some of the radiative transfer calculations were done specifically for this work (see below). The simulations were performed using a modified version of the \textsc{gadget-2} cosmological $N$-body/smoothed particle hydrodynamics (SPH) code \citep{Springel2001}. Every $10-100$ Myr, the simulation outputs were post-processed using the \textsc{sunrise} \citep{Jonsson2006,Jonsson2010} dust radiative transfer code to yield SEDs of the simulated galaxies for $7$ isotropically distributed camera angles. The success of this approach at reproducing SEDs characteristic of typical star-forming galaxies \citep{Lanz2014}, $z \sim 2$ dusty star-forming galaxies \citep{Narayanan2010a,Narayanan2010,Hayward2012}, and AGN \citep{Snyder2013} make it a natural choice for comparison with our observed sample. Further details regarding \textsc{gadget} and \textsc{sunrise} and the specific simulation library that we use are given in the subsequent subsections.

\subsection{Hydrodynamical Simulations} \label{sec:gadget}

\textsc{gadget-2} \citep{Springel2005gadget} computes gravitational forces using a tree-based gravity solver. Hydrodynamics is treated using {\rcom a modified} TreeSPH \citep{Hernquist1989}, in a fully conservative manner \citep{Springel2002}. The simulations include radiative heating and cooling following \citet{Katz1996}. Star formation is modeled by applying the volume-density-dependent Kennicutt-Schmidt relation $\rho_{\rm SFR} \sim \rho_{\rm gas}^{1.5}$ \citep{Kennicutt1998a} with a density cutoff at $n \sim 0.1 \ \rm{cm^{-3}}$ on a particle-by-particle basis. {\rcom This {\rcomtwo normalization of this} prescription is tuned to reproduce the galaxy scale KS relation.} 
{\rcomtwo We adopt an effective equation of state following} the two-phase subresolution ISM model of \citet{Springel2003}, which accounts for the effects of supernovae feedback in the form of heating and the evaporation of gas \citep{Cox2006b}, is used. 
Explicit stellar winds are not included in our simulations. Each gas particle is self-enriched according to its SFR, assuming a yield of $y \sim 0.02$. We employ the black hole accretion and feedback model of \citet{Springel2005feedback}. Each galaxy is initialized with a black hole sink particle with initial mass $10^5 \textrm{M}_{\odot}$ that accretes at the Eddington-limited Bondi-Hoyle rate and {\rcomtwo 5\% of the luminous energy of the AGN is returned to the ISM as feedback in the form of thermal energy}, to match the $M-\sigma$ relation \citep{DiMatteo2005}. The simulations adopt a standard 10\% {\rcomtwo radiative} efficiency such that the AGN luminosity is given by $L_{\rm{bol}} = 0.1 \dot{M}_{\rm{BH}} c^2$ \citep{Shakura1973}. For additional and more detailed information concerning \textsc{gadget}, see \citet{Springel2005gadget}.

\subsection{Radiative Transfer Calculations} \label{sec:sunrise}

To calculate UV--mm SEDs of the simulated galaxies, we perform dust radiative transfer in post-processing using the 3D Monte Carlo radiative transfer code \textsc{sunrise}, 
which calculates how emission from stellar and AGN particles in the \textsc{gadget-2} simulations is absorbed, scattered, and re-emitted by dust. Star particles from the \textsc{gadget-2} simulations are treated as single-age stellar populations. Those aged $>10 \ \rm{Myr}$ are assigned \textsc{starburst99} \citep{Leitherer1999} template SEDs according to their ages and metallicities, whereas those with ages $<10 \ \rm{Myr}$ are assigned templates from \cite{Groves2008}, which include emission from H\textsc{ii} and photodissociation regions (PDRs) surrounding the clusters. Black hole particles, with luminosity defined in Section~\ref{sec:gadget}, are assigned the luminosity-dependent AGN SED templates of \cite[][hereafter H07]{Hopkins2007}, which are empirical templates based on observations of unreddened quasars. In the IR, these templates match the mean quasar SED of \citet{Richards2006a}.  
Once the spatial distribution and SEDs of sources (i.e. stars and AGN) are specified, the dust density field must be determined. To do so, the \textsc{gadget} gas-phase metal density is projected onto a 3D octree grid initially $200 \ \rm{kpc}$ on a side. To calculate the dust density, it is assumed that $40 \%$ of the metals are in the form of dust \citep{Dwek1998}. The default dust model is the Milky Way model of \citet{DL07}. Grid cells are refined until both the variation in the metal density within a grid cell and the total optical depth through a grid cell are less than specified thresholds or until a maximum number of refinement levels is reached; see \citet{Jonsson2010} for details. We have confirmed that the grid refinement parameters ensure that the SEDs are converged within $\sim 10 \%$ \citep{Hayward2011}.

After the above steps, we propagate $10^7$ photon packets through the grid to calculate how the stellar and AGN emission are absorbed and scattered by dust. Then, the radiation absorbed by dust is re-emitted, assuming that the large grains are in thermal equilibrium. A fraction of the small dust grains are assumed to emit thermally, whereas the rest emit an empirically based PAH template \citep{Groves2008}. This fraction is fixed to 50$\%$ following \citet{Jonsson2010} to match mid-IR flux ratios from SINGS \citep{Dale2007}\footnote{The question of how much mid-IR continuum is assigned to SF is uncertain, and complicates a direct comparison with our empirical $f{\rm (AGN)}_{\rm MIR}$ fractions.  The extent of which is beyond the scope of this discussion. In this paper we focus on total IR AGN fractions where this effect is small.}. 
The IR emission is then propagated through the grid to account for dust self-absorption, and the equilibrium dust temperatures are recalculated. This process is iterated until convergence. The final result of the radiative transfer calculation is spatially resolved far-UV--mm SEDs (i.e. integral field spectrograph-like data) of the simulated galaxy seen at different times (every 10-50 Myr for the simulations used in this work) from 7 viewing angles. For the purposes of this work, we sum the SEDs of each pixel to yield the integrated SED of the system.

The default behavior of \textsc{sunrise} is to calculate the AGN luminosity using the black hole accretion rate from the \textsc{gadget-2} simulation and assuming a standard radiative efficiency of 10\%; we denote these runs as AGN1$\times$. For this work, as in \citet{Snyder2013}, we also performed radiative transfer calculations in which we assumed radiative efficiencies of 100\% (AGN10$\times$) or 0\% (AGN0$\times$). This simulates the effects of short-term stochasticity in the black hole accretion rate \citep[e.g.][]{Hickox2014} that is not present in the time-averaged accretion rates from the \textsc{gadget-2} simulation snapshots (although the accretion rates corresponding to individual timesteps exhibit considerable variation; \citealt{Hayward2014arepo}). Moreover, by computing the radiative transfer with the AGN emission disabled, we are able to directly disentangle the effect of AGN emission on the resulting SED. Note that in all cases, the same hydrodynamical simulation is used; i.e. thermal AGN feedback is included assuming a radiative efficiency of 10\% \citep[for details see][]{Snyder2013}. 

\subsubsection{Host Galaxy ISM Treatment \label{sec:ism}}

The ISM treatment used in the hydrodynamic simulations is the two-phase model of \citet{Springel2003} (see Section~\ref{sec:gadget}). 
{\rcom Each resolution element is implicitly assumed to contain a warm ($>10^5$\ K) and cold ($<10^4$\ K) gas component, but only a single density and an ``effective pressure" is actually evolved.}
How the sub-resolution ISM structure is treated can affect the resulting SED significantly, as discussed in detail in various previous works \citep[e.g.][]{Younger2009,Hayward2011,Snyder2013,Lanz2014}. To crudely capture the uncertainty caused by not resolving the full structure of the ISM, we use two extreme cases when performing the radiative transfer. The `multiphase off' treatment uses the total dust density ({\rcom spreading the total dust mass uniformly through the cell}) to calculate the optical depth of each cell; this yields an upper limit on the optical depth through a cell. `Multiphase on' {\rcom assumes that the unresolved cold clouds} have a negligible volume filling factor and {\rcom and thus removes the dust} implicitly contained {\rcom in this phase (according to the sub-grid model)} when performing the radiative transfer. This yields a lower limit on the optical depth through a cell. The effect of each assumption on the mid-IR AGN spectral signatures is discussed in \citet{Snyder2013}, who generally find their results are not significantly dependent on the model adopted. We examine how our results depend on the sub-resolution ISM treatment in Appendix\,\ref{sec:agnism}.

\begin{deluxetable}{cccl}
\tabletypesize{\footnotesize}
\tablecolumns{4}
\tablewidth{\columnwidth}
\tablecaption{Model Progenitor Initial Properties \label{tab:modeltable}}
\tablehead{ 
\colhead{Progenitor} & \colhead{$M_{\rm{*,init}}$} & \colhead{$f_{\rm{gas,init}}$} & \colhead{Reference}
\\
\colhead{Name} & \colhead{($h^{-1} \ \rm{M_{\odot}}$)} & \colhead{} & \colhead{Name\tablenotemark{a}}
}
\startdata
M1 & $3.78 \times  10^{9}$ & $0.26$ &  M1[J10,L14]  \\
M2 & $1.18 \times 10^{10}$ & $0.21$ & M2[J10,L14]  \\
M3 & $4.23 \times 10^{10}$ & $0.16$ &  M3[J10,L14] \\
M4 & $3.39 \times 10^{10}$ & $0.40$ & vc3[S13,H15] \\
M5 & $4.08 \times 10^{10}$ & $0.60$ & c5[H12]  \\
M6 & $1.56 \times 10^{10}$ & $0.60$ & c6[H11,H12,S13]  \\
M7 & $2.08 \times 10^{10}$ & $0.80$ & b5[H12]  \\
M8 & $8.00 \times 10^{10}$ & $0.80$ & b6[H12] 
\vspace{-0.2cm}
\enddata
\vspace{-0.2cm}
\tablenotetext{a}{Name in the literature. References are: \citep[J10;][]{Jonsson2010}, \citep[H11;][]{Hayward2011},\citep[H12;][]{Hayward2012},\citep[S13;][]{Snyder2013}, \citep[L14;][]{Lanz2014}, \citep[H15;][]{Hayward2015}. }
\end{deluxetable}
\vspace{-0.6cm}
\subsection{Simulation Library} \label{sec:simlib}
Table~\ref{tab:modeltable} shows the 8 progenitors in our simulation library. These progenitors are simulated both as isolated disks and identical mergers. Mergers are particularly relevant because it is believed that gas inflows during such events are a primary trigger for exciting bright IR activity \citep{Barnes1992, Barnes1996, Mihos1996, Hopkins2010b}.  All mergers use the {\rcomtwo tilted prograde-prograde} `e' orbit from \citet{Cox2006a} (with the exception of \vc{}, which uses the {\rcomtwo retrograde-retrograde} `c' orbit).  In addition, to the 8 mergers performed with the default AGN strength (i.e., assuming a radiative efficiency of 10\% to determine the AGN luminosity), we also ran no-AGN variants of the radiative transfer calculations on all mergers excluding M5 and M7\footnote{{\rcom As seen in Table~\ref{tab:modeltable}, models M5 and M7 do not occupy drastically different parameter spaces than M6 and M8, respectively. This omission should not significantly affect the conclusions of this paper.}}.
This enables us to directly determine the effect of the AGN on the UV--mm SED. For most mergers, again excluding M5 and M7, we also performed AGN10$\times$ calculations (i.e., assuming a radiative efficiency of 100\%). 
The AGN0$\times$ calculations are not used to fit the observed SEDs, but are a means of calculating the effect of the AGN on the emergent UV--mm SED, post host galaxy dust reprocessing of the AGN light i.e., $L_{\rm{AGN,reprocessed}} = L_{\rm{total,AGN1\times}} - L_{\rm{total,AGN0\times}}$. Including the isolated disks, AGN1\,$\times$, and AGN10\,$\times$ mergers, our observed SEDs are fit to a suite of 22 simulations.
 
The simulations most analogous to $z \sim 0$ galaxies (M1-M4, Table~\ref{tab:modeltable}) are prescribed the `multiphase on' ISM treatment (Section~\ref{sec:ism}) by default. Using a restricted sample, we found that using the `multiphase off' ISM treatment did not significantly alter the goodness-of-fit values obtained. 
For the more luminous and gas-rich $z \sim 3$ simulations (M5-M8), the `multiphase off' treatment (Section~\ref{sec:ism}) was used because this treatment was found to yield better agreement with the SEDs of high-redshift dusty star-forming galaxies (see \citealt{Hayward2011} for discussion). Both ISM assumptions are only approximations, and the truth should lie somewhere in between. The need for these assumptions could potentially be eliminated (or at least mitigated) via use of state-of-the-art simulations that resolve the structure of the ISM \citep{Hopkins2013, Hopkins2014}; this is a focus of ongoing work. In Appendix~\ref{sec:agnism}, we examine the potential effects of the ISM treatment on our results and find that they are negligible.
 
Other factors likely to affect the emergent SED include variations in merger orbital parameters, the adopted AGN template, and the dust model. In Appendix~\ref{sec:agntemp}, we investigate how our results depend on the choice of AGN template \citep[see also discussion in ][]{Snyder2013}. In Paper III (Roebuck et al. in prep.) we explore such systematics further, including run these simulations with different AGN templates, different dust compositions (MW, SMC, and LMC) and different ISM treatments. The spread in the emergent SED allows us to estimate a model uncertainty which we adopt in our SED fitting (Section~\ref{sec:fitting}). 

\section{Analysis} \label{sec:analysis}

\subsection{AGN Fractions from Our Simulations} \label{sec:agnfracs}

\begin{figure*}
   \centering
   \includegraphics[width=1.95\columnwidth]{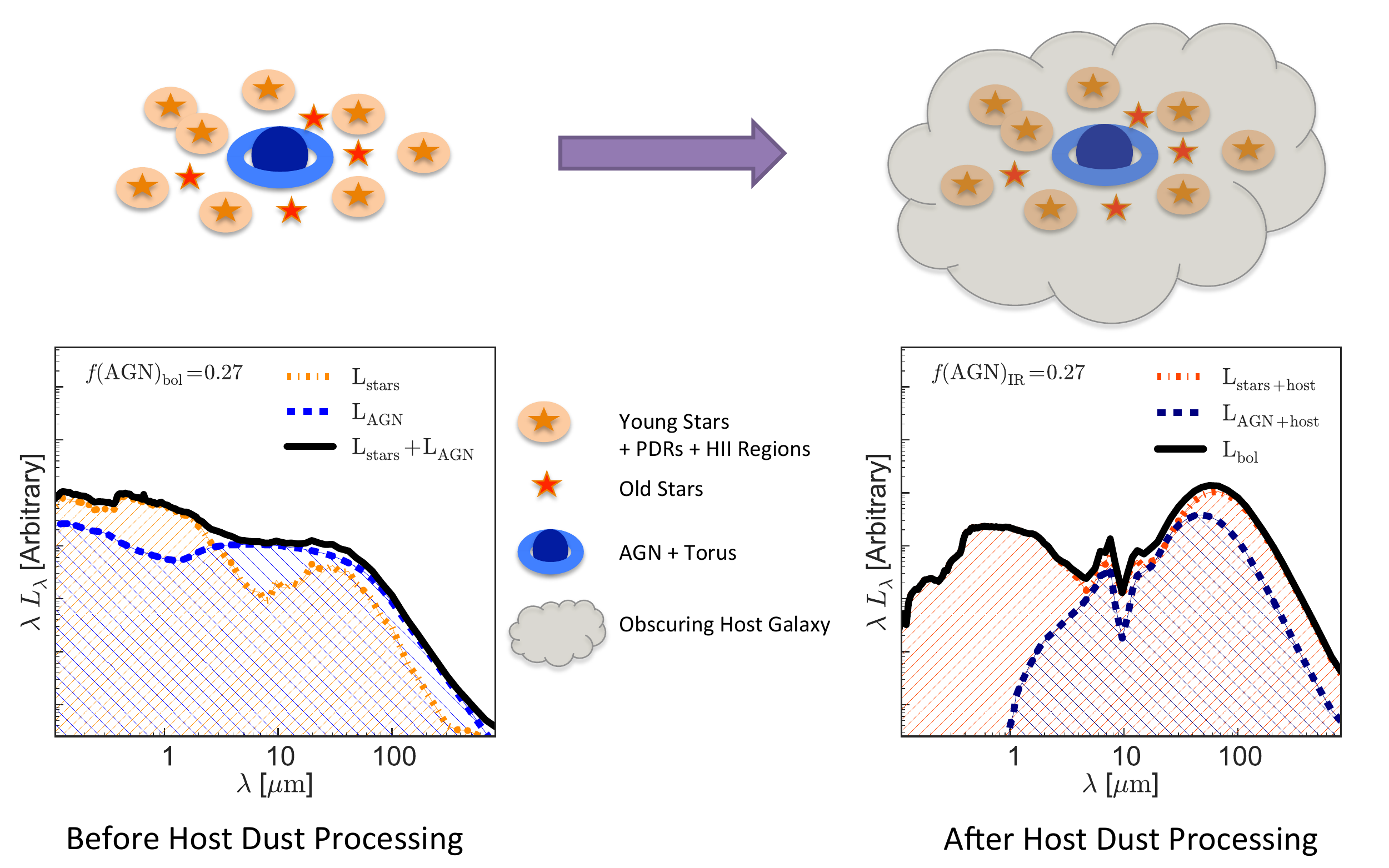} 
   \caption{
  A schematic view of the stellar and AGN SEDs in our simulations before (\emph{left}) and after (\emph{right}) host galaxy dust reprocessing. For this example, we used a gas-rich merger of M6 (Table~\ref{tab:modeltable}) at coalescence. The top panels illustrate that the stellar component includes both old stars (small red stars) in addition to young stars and their associated H\textsc{II}+PDR regions (orange stars with small halos). The AGN component includes emission from its dusty torus as well as likely some NLR emission (shown as the black hole and blue torus). The bottom panels show the stellar SED (orange curve), the AGN SED (blue curve), and the total SED (black curve). Comparing the left and right-hand bottom panels shows that both the stellar and AGN components are strongly reprocessed by the dust in the simulated host galaxy; as a result, the AGN contributes significantly to the far-IR SED peak. 
 }
   \label{fig:sedpicture}
\end{figure*}

\begin{figure*}[h]
   \centering
   \includegraphics[width=1.95\columnwidth]{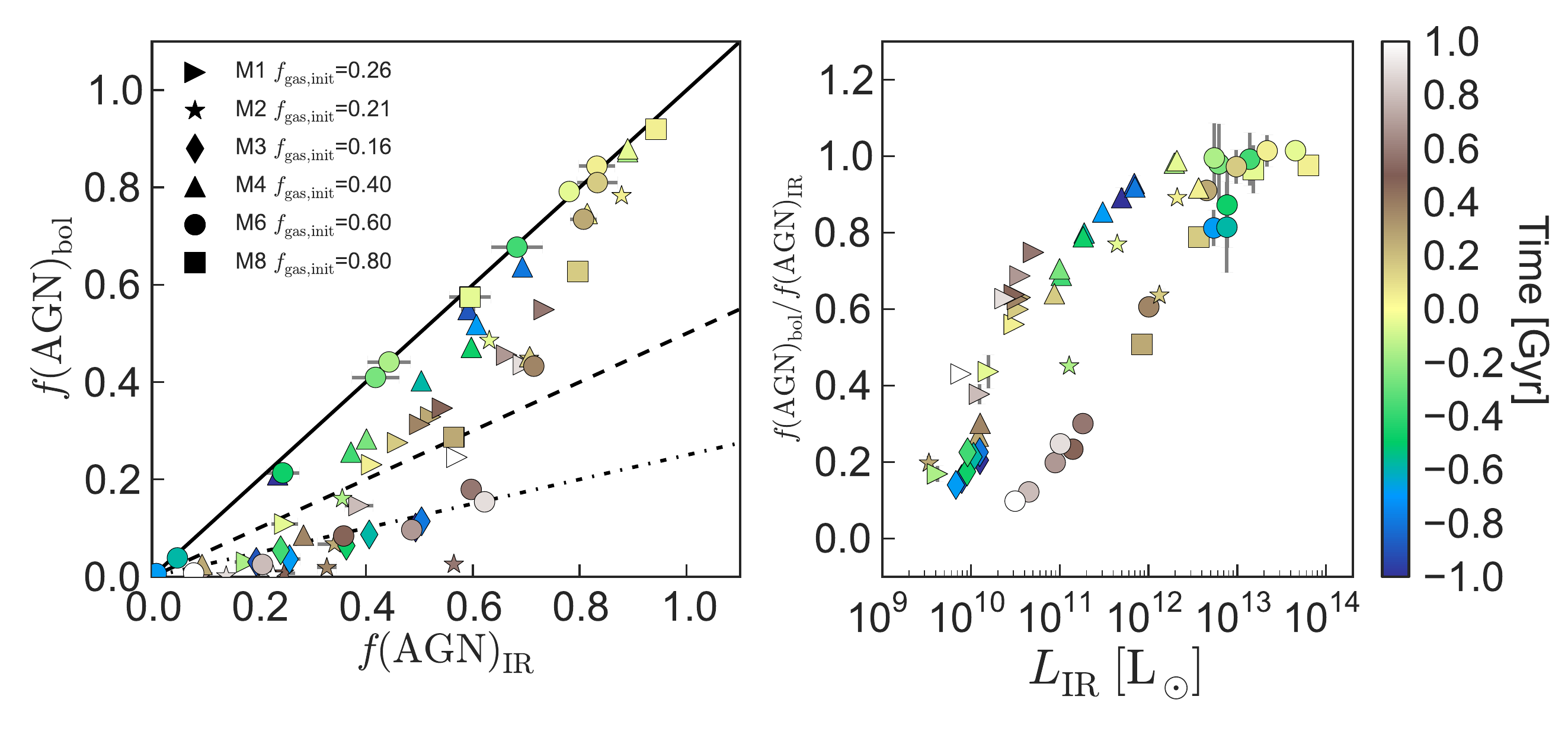} 
   \caption{{\it Left:} Our simulated AGN fractions (defined in Section~\ref{sec:agnfracs}) for all AGN10$\times$ merger simulations (see Table~\ref{tab:modeltable}). 
   {\rcom The lines show \inp{} $= \alpha \times \hosteqn$ where $\alpha=1$ (solid), $=1/2$ (dashed), and $=1/4$ (dot-dashed).}
   {\it Right:} The ratio of the two fractions ($R$= \inp{} / \host{}, see Equation\,\ref{eq:ratio_def}) vs. luminosity. To avoid cluttering this figure, and to accommodate varying time coverage and resolutions across our library, the points are interpolated onto a common array with $\Delta t\sim100$\ Myrs, but we tested that our conclusions are unaffected by the choice of time bin here. Each point is the average of the 7 viewing angles, with the error bars representing the standard deviation thereof. The symbols are colored relative to coalescence (here time=0.0) as indicated by the color-bar. 
   This figure shows that the two fractions agree best for the highest initial gas fraction, and consequently reaching highest luminosity systems for the times up to and including coalescence. Post-coalescence, $R$ drops suggesting the IR AGN fraction is no longer a good proxy for the total AGN fraction. 
   }
   \label{fig:agncollage}
\end{figure*}

\begin{figure*}[h]
   \centering
   \includegraphics[width=1.95\columnwidth]{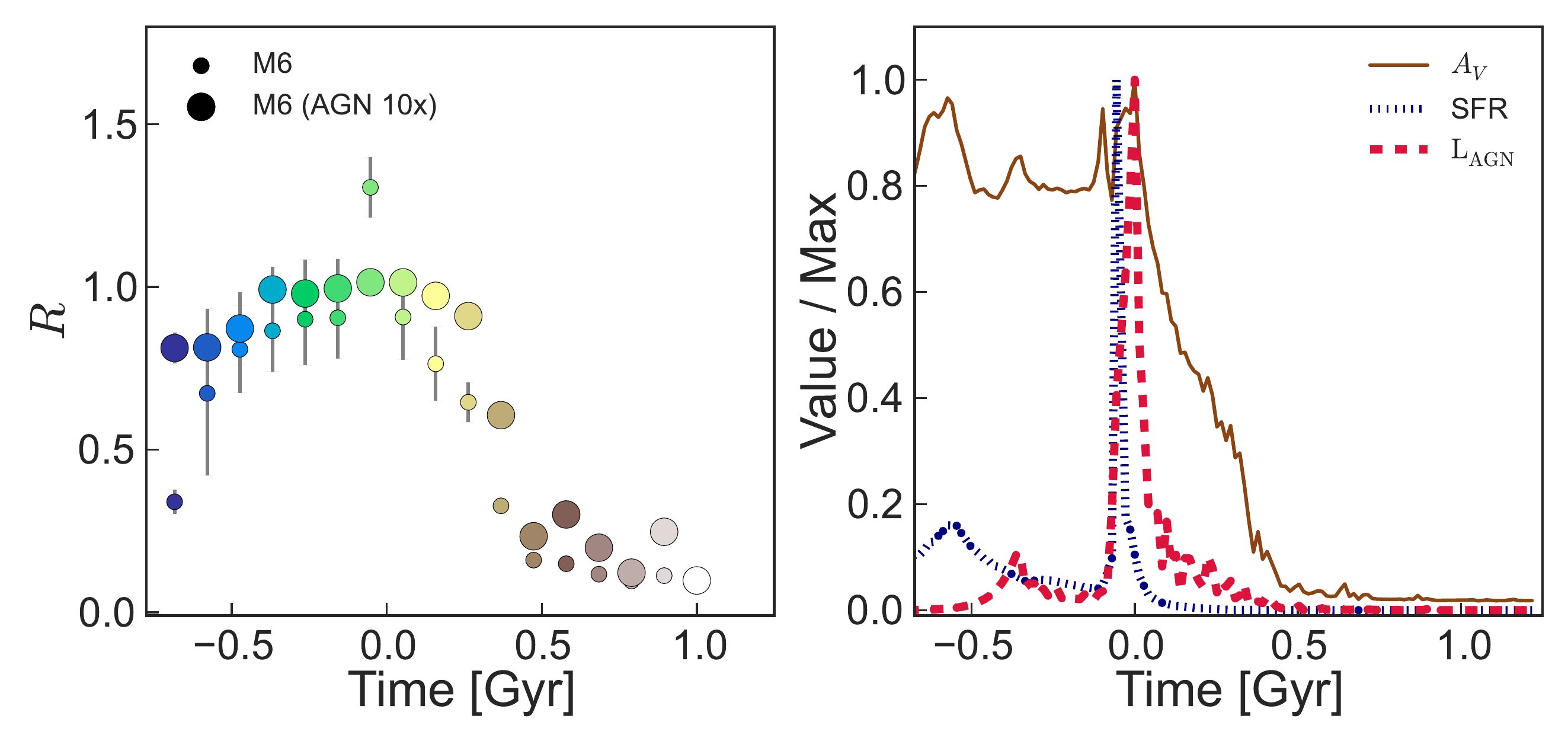} 
   \caption{
   \emph{Left:} The ratio of bolometric to IR AGN fractions ($R$, see Section~\ref{eq:ratio_def}) as a function of time for the fiducial (small circles) and AGN10$\times$ (large circles) mergers of model M6. This ratio is $\sim$1 until just after coalescence. \emph{Right:} The evolution of $A_V$, SFR, and $L_{\rm AGN}$ for the fiducial (AGN1$\times$) merger of model M6, as a function of time all normalized by their maximum values. Post-coalescence, the rate of decrease of $R$ with time most closely mimics the rate of decrease in $A_V$ suggesting that $A_V$ may be a good observable indicator of how closely the IR AGN fraction traces the total fraction.
   }
   \label{fig:fagnratio}
\end{figure*}

\begin{figure}[h]
   \centering
   \includegraphics[width=.95\columnwidth]{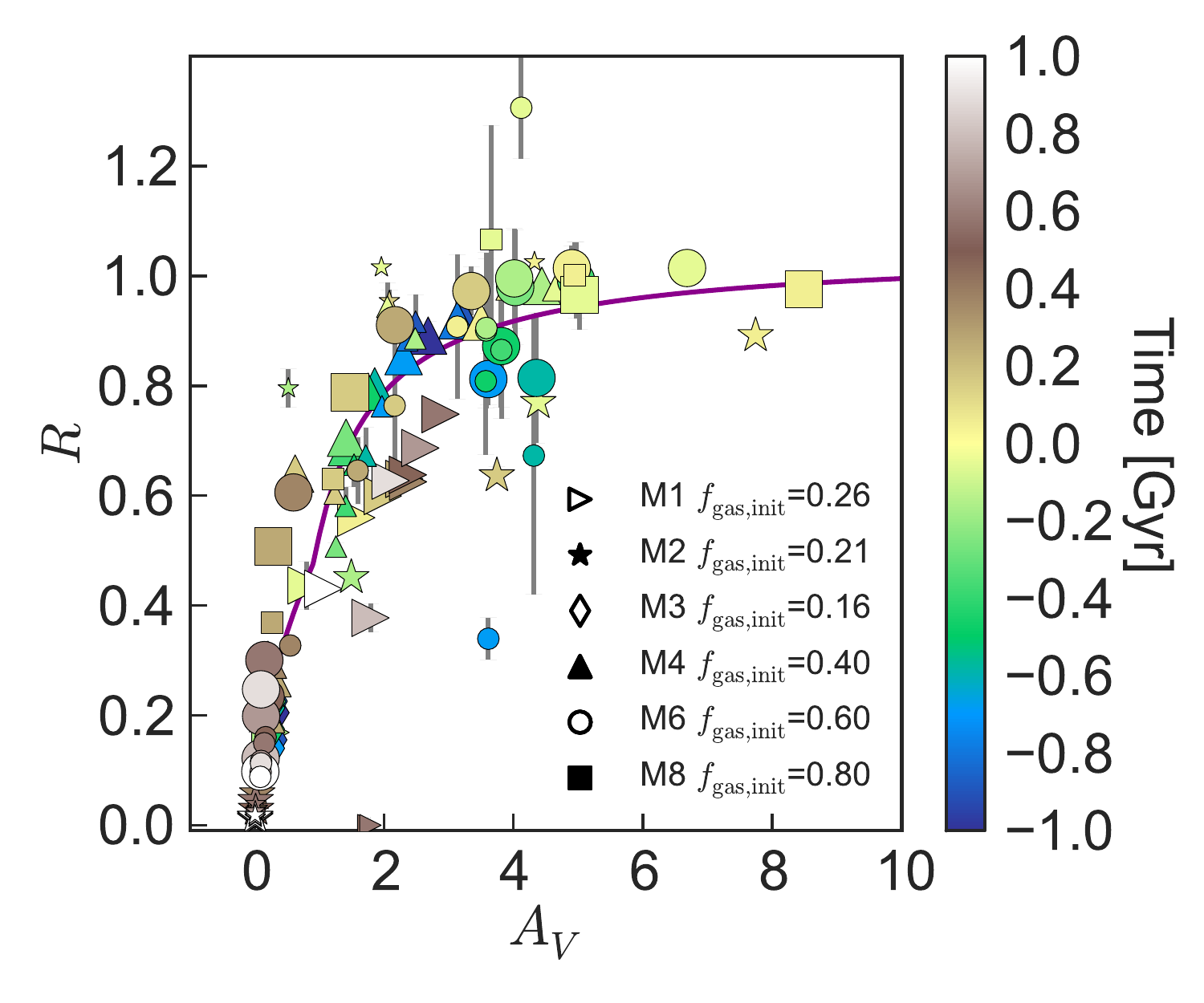}
   \caption[width=0.95\columnwidth]{The ratio of the bolometric IR AGN fractions ($R$, see Equation\,\ref{eq:ratio_def}) as a function of {\rcom $A_V$(=$-2.5 \log L_{V} /  L_{V, {\rm input}}$)}. As in Figure~\ref{fig:agncollage}, the colors are time relative to coalescence, the time sampling is in steps of 100\,Myrs, the data points are the mean values over viewing angle, and the error bars are the standard deviation thereof. We show the full range in initial gas fractions and both fiducial (AGN1$\times$) and boosted (AGN10$\times$) models as respectively the smaller and larger symbols. Overlaid is the best fit relation given in Equation~\ref{eq:fagnratio}. This shows that, largely independent of initial conditions, as $A_V$ grows larger $R$ approaches unity and vice versa. Thus IR AGN fraction measurements are useful measures of bolometric AGN strength when the extinction is high. 
   }
   \label{fig:fagnvav}
\end{figure}

The first question we want to address with our simulations is what is the total contribution of the AGN to the IR emission of our simulated galaxies -- including not only the IR emission associate with the nuclear regions of the galaxy (i.e. from the torus), but also dust in the host galaxy heated by the AGN. Figure~\ref{fig:sedpicture} helps to visualize the role of the ISM in the host galaxy in re-distributing the emission from both the AGN and the stars{\rcom ,} where the left-hand panel shows the SEDs of stars and AGN before host galaxy processing, and the right-hand panel shows the same post host galaxy processing. 
The AGN SED post-processing is significantly redder (more IR heavy) compared to the input AGN SED pre-processing{\rcom ,} highlighting the importance of accounting for AGN heating of the host galaxy dust. In this paper, we do not investigate the spatial extent of the AGN-powered FIR emission; this will be discussed in Hayward et al. (in prep.). 
To determine the total post host dust processing IR emission that results from the AGN, we perform the radiative transfer calculations both with and without the AGN emission and then take the difference between the two SEDs. This is given in Equation\,\ref{eq:agn_host} below. 
\vspace{-0.3cm}
\begin{equation}
\centering
\hosteqn=\left(\frac{L_{\rm{total}}  - L_{\rm total,no AGN} }{ L_{\rm{total}} } \right)_{8-1000 \rm{\mu m}}
\label{eq:agn_host}
\end{equation}
where $L_{\rm{total}}$ is the post-radiative transfer SED including both the attenuated stellar and AGN emission in addition to dust emission (the fiducial calculation), and $L_{\rm{total,no AGN}}$ is the post-radiative transfer SED for the calculation in which the AGN emission is ignored (the red-orange curve in the right panel of Figure~\ref{fig:sedpicture}; see Section~\ref{sec:simlib} for details). The role of the input AGN SED template assumed is addressed in Appendix \ref{sec:agntemp}. The key conclusion thereof, is that $\hosteqn$ is not strongly dependent on the input AGN SED. This conclusion is due to the spatially and spectrally integrated nature of this AGN fraction.

The second key question is, assuming we have perfect knowledge of the AGN contribution to the IR, is this a good measure of the intrinsic AGN fraction? 
We obtain this by dividing the integrated AGN luminosity in the 0.1-1000\um\ regime over the integrated total (AGN+stars) luminosity (see left-hand panel in Figure~\ref{fig:sedpicture}). {\rcom While this fraction is missing the X-ray and radio emission, for simplicity, we still refer to it as the \emph{bolometric or ``bol" AGN fraction}}. This is defined in Equation\,\ref{eq:agnintrinsic} below. 
\begin{equation}
\centering
\fagnbol= \left( \frac{ L_{\rm{AGN}} }{ L_{\rm{stars}} + L_{\rm{AGN}} } \right)_{0.1-1000 \rm{\mu m}} 
\label{eq:agnintrinsic}
\end{equation}
where $L_{\rm{stars}}$ is the integrated UV-mm luminosity of the emission from stars, PDRs and H\textsc{ii} regions (the latter two are powered by star formation, so the luminosity would be the same if we considered the intrinsic stellar emission before it is processed in the PDRs and H\textsc{ii} regions). 

Because we calculate it using the intrinsic SEDs, {\rcom $\fagnbol$} is independent of viewing angle. If we were to calculate it from the post-radiative transfer SEDs, the value inferred from a given line of sight would differ from the intrinsic value, but the intrinsic value would be recovered if we averaged over sufficiently many viewing angles. We remind the reader that the AGN luminosity is calibrated by the accretion rate (see Section~\ref{sec:gadget}) for $L_{\rm AGN,bol}$ from X-ray through mm, and that $\lesssim10\%$ of the AGN flux is emitted in the X-ray, 0.5-10\,keV, regime \citep{Hopkins2007,Snyder2013}. Lastly, we define the ratio of the above AGN fractions as:

\begin{equation}
\centering
R=\frac{\fagnbol }{\hosteqn }
\label{eq:ratio_def}
\end{equation}

 \subsection{Trends with Merger Stage} \label{sec:role_mergerstage}
 
Figure~\ref{fig:agncollage} {\it left} shows the relationship between our AGN fractions (Section~\ref{sec:agnfracs}), as a function of time for six AGN10$\times$ merger simulations (Section~\ref{sec:simlib}) which have different initial gas fractions, as indicated. We chose the boosted models because they reach higher AGN fractions making trends easier to see, but as we discuss in the following Section, our final conclusions are valid for both fiducial (AGN1$\times$) and boosted models. 

The time relative to coalescence, defined as the moment the black hole separation goes to zero, is indicated by the colorbar located in the upper-right panel of the figure. Overall, the higher gas fraction models achieve higher AGN fractions peaking around coalescence when the two fractions are on the 1:1 relation. Lower gas fraction models achieve overall lower AGN fractions, and also fall somewhat short of the 1:1 relation even at coalescence. With these differences, in mind, all models show reasonably good agreement between the two fractions up to and including coalescence, but show much higher $\hosteqn$ vs. $\fagnbol$ post-coalescence. In this regime, the AGN fractions are overall much lower.

Figure~\ref{fig:agncollage} {\it right} shows the ratio of $\hosteqn$ to $\fagnbol$ vs IR luminosity. The two tracks seen in this figure correspond to different initial gas fractions (upper track is for models with $f_{\rm gas,init}<0.5$ whereas the lower track is for models with $f_{\rm gas,init}>0.5$). Therefore, for all models, the higher the IR luminosity, the higher the $R$ parameter (i.e. the more closely IR AGN fractions trace UV-mm AGN fractions).  For any given model, the highest luminosities are achieved around coalescence. However, for a fixed IR luminosity, different initial gas fraction models will correspond to different stages in the merger and will have different $R$ values. For example, a $10^{11}$\lsun\ model galaxy can have $R$ values that differ by a factor of 3 (0.2-0.6), with the higher values corresponding to lower initial gas fraction models nearer coalescence.  This degeneracy is lifted for the highest luminosities ($>10^{12}$\lsun) where all models correspond to high $R$ values. Aside from lensed systems (e.g. \citet{Sklias2014}), $L_{\rm IR} \gtrsim 10^{11}$\lsun\ corresponds to the sensitivity limit for current wide field IR extragalactic surveys. 

Because initial gas fraction or merger stage are difficult to determine observationally, the above finding makes it difficult to know, apart from the highest luminosity sources, whether or not the IR AGN fraction for a given galaxy is a good proxy for the overall AGN fraction or not. 

To try and break this degeneracy observationally, Figure~\ref{fig:fagnratio} examines more closely the reasons behind the above trends. Here we explicitly show $R$ as a function of time for the gas-rich merger M6 both in the AGN1$\times$ and 10$\times$ case. The right-hand panel shows the time evolution of the SFR, $L_{\rm AGN}$, and $A_V$. 
{\rcom By contrast $R$ (as well as $A_V$) are relatively flat until post-coalescence. 
Therefore $A_V$ most closely traces the evolution of $R$ with time. }
For clarity, we only show one model in Figure~\ref{fig:fagnratio}, but the trends are qualitatively the same for all models except that the models with lower $f_{\rm gas,init}$ reach lower maximal $A_V$ values.  
{\rcom Figure~\ref{fig:fagnratio} shows that $R$ is not a good proxy for the AGN luminosity which is much more peaked around coalescence. In addition, we found no clear trends between either the bolometric or IR AGN fractions and $R$.} 

In the following section we explore further this dependence on $A_V$ as well as provide a relation for converting the IR AGN fraction to a bolometric (here UV-mm) AGN fraction. 

\subsection{Trends with $A_V$} \label{sec:interp_av}

Figure~\ref{fig:fagnvav} shows explicitly the universality of this dependence on $A_V$. Here we show $R$ as a function of $A_V$ for all models including default (AGN1$\times$) and boosted (AGN10$\times$) cases. As in Figure~\ref{fig:agncollage}-\ref{fig:fagnratio}, the points are averaged over camera angle with the errorbars showing the standard deviation thereof. We find that $R$ is $\sim$\,1 for the most obscured models and timesteps ($A_V \gtrsim$ 3), which suggests that while the optical depth of the host galaxy is high the IR AGN fraction does trace the inherent AGN to stellar strength. This ratio remains $>$\,0.5 (i.e. the IR AGN fraction traces the intrinsic UV-mm AGN fraction within a factor of 2) for all cases when the dust extinction is high (i.e. $A_V \gtrsim$ 1). The spread between the models is roughly consistent with the spread between camera angles. The only outlier is a point at $A_V$\,$\sim$\,3.9 which corresponds to the initial few timesteps in the M6 default (AGN1$\times$) run (see Figure~\ref{fig:fagnratio}). In this regime, the AGN luminosity is negligibly small, with both AGN fractions tiny. Where there might be a true physical effect, it is likely that this outlier is noise related to the very small numbers. Other models that cover this regime (the boosted M6 and the boosted and unboosted M4), do not show this behavior {\rcomtwo supporting the view that} it is likely an outlier. Still, this might be a behavior associated with more isolated systems. As none of our isolated disk models have the AGN0$\times$ runs necessary for explicit determination of $R$, we do not investigate this further in the present paper.

{\rcom
The relation between $R$ and $A_V$ is approximated by:
}

\begin{equation}
R = \left\{
        \begin{array}{ll}
            \frac{-0.52}{A_V} + 1.05 & \quad A_V \geq 1 \\
            0.51 \sqrt{A_V} - 0.01 & \quad A_V < 1
        \end{array}
    \right.
    \label{eq:fagnratio}
\end{equation}
{\rcom
This piecewise form qualitatively represents the trend in Figure~\ref{fig:fagnvav}, however we do not compute a formal fit as the points shown in Figure~\ref{fig:fagnvav} are binned in time and averaged by viewing perspective for clarity. We tested that this trend is maintained with the raw data plotted as well.
}

But why should $R$ drop at lower $A_V$ (typically post-coalescence)? Post-coalescence, the AGN luminosity as well as the SFR both decrease dramatically. However, thanks to the recently build-up stars, the bolometric luminosity is strongly dominated by the now aging stellar population \citep[see also][]{Donoso2012,Hayward2014}. The stellar SED here has relatively little UV emission compared to optical/near-IR emission. By contrast, the assumed constant AGN SED shape is rich in UV photons even in the post-coalescence regime (although see Section\,\ref{sec:discussion} for caveats). This allows the galaxy to have a higher IR than UV-mm AGN fraction (since UV photons are much more efficiently absorbed than optical photons). Therefore post-coalescence the boost in the optical-NIR coming from the older stellar populations, coupled with the redder stellar SED relative to the AGN, lead to lower $R$ values in this regime. We reiterate that this does not imply high values of $\hosteqn$ post coalescence compared to earlier times, as $\hosteqn$ is typically much lower post coalescence (see Figure~\ref{fig:agncollage}). 

In addition, differential extinction between the AGN and the stars can also contribute to this effect. In essence, while the attenuation of the starlight is given by $A_V$ (which dominates the visible light), there might be greater column of dust toward the AGN. 
{\rcom {\rcomtwo The ISM of these simulated galaxies is smoother} than real ones and thus harder to get clear sight lines to an AGN in the presence of significant host dust than in reality. 
\footnote{{\rcom The } \textsc{sunrise} {\rcom calculations may not result in unobscured AGN because all light-emitting particles are `fuzzy'; photons start from a random position within a sphere with radius equal to the gravitational softening, which is significantly larger than the scale of the torus emission. This has the effect of `smearing out' the emission, which leads to less variation in the effective optical depth than you would get if it were calculated for individual lines of sight. It is only when you allow for a clumpy {\rcomtwo ISM on scales below the resolution limit} that you can get low-attenuation lines of sight (discussed in \citet{Hopkins2012}).}}
The error bar on R in Figure~\ref{fig:fagnvav} represents the spread across the 7 sight lines we track, but this spread is likely an underestimate of the the spread we would get in real galaxies given the smoothness of the simulated galaxies. 

Indeed, we do not observe cases of significant AGN luminosity when the galaxy is not dusty.
{\rcomtwo Thus Type-1 QSOs} are missing in our model library. Because of selection bias \citep[see][]{Kirkpatrick2015} they are missing from our observed sample as well. }
We examine this issue more closely in upcoming papers (Roebuck et al. in prep, and Hayward et al. in prep.)

We caution, \citet{Hayward2015} find the $A_V$ inferred via SED modeling with \textsc{magphys} \citep{daCunha2008} tends to be less than the true $A_V$ when $A_V \gtrsim$1. When the attenuation is high, the only UV-optical emission originates from the relatively unobscured stars. Thus $A_V$ appears lower than it actually is. This underestimation is most severe for merger simulations near coalescence, however the difference is typically small ($<$0.2 in $A_V$, with overall less extreme systems with $A_V\sim2$.). 
Therefore, observationally determined $A_V$ would be a lower limit to the true ones resulting in a lower limit on $R$ if Equation\,\ref{eq:fagnratio} is used. Because $R$ as a function of $A_V$ flattens for high $A_V$, the effect of this in practice is minimal. 

\begin{deluxetable*}{ccccccc}
\tabletypesize{\footnotesize}
\tablecolumns{7}
\tablewidth{2\columnwidth}
\tablecaption{Median Best Fit Parameters \tablenotemark{a} \label{tab:bestfitclass}}
\tablehead{ 
\colhead{Class} & \colhead{$N_{\rm{gal}}$} & \colhead{$\chi^2$} & 
\colhead{$A_V$} &
\colhead{\inp{}} & 
\colhead{\host{}}\tablenotemark{b}  &
\colhead{\tir{}$_{\rm{,emp}}$}\tablenotemark{c} 
}
\startdata
SFG & $         101$ & $  0.88 \pm   0.59$ & $  3.79 \pm   0.97$ & $ 4 \pm  10$ & $ 6 \pm  23$ ($ 4 \pm  11$) & $ 0 \pm  0$ \\
Composite & $         116$ & $  0.97 \pm   0.63$ & $  3.87 \pm   1.06$ & $ 38 \pm  20$ & $ 52 \pm  17$ ($ 41 \pm  22$) & $ 15 \pm  10$ \\
AGN & $         119$ & $  1.30 \pm   0.76$ & $  3.87 \pm   1.25$ & $ 66 \pm  18$ & $ 78 \pm  15$ ($ 77 \pm  17$) & $ 55 \pm  7$ 
\enddata
\tablenotetext{a}{The uncertainties given are the median absolute deviation among all galaxies of the given class.}
\tablenotetext{b}{For simulations with AGN$0\times$ runs, this is calculated from Equation~\ref{eq:agn_host} ($\sim$ 60\% of the sample; the first number); otherwise, it is calculated from Equation~\ref{eq:fagnratio}. The number in brackets includes all sources.}
\tablenotetext{c}{Empirical values derived in \citet{Kirkpatrick2015}.}
\end{deluxetable*}

\begin{figure*}
   \centering
   \includegraphics[scale=0.38]{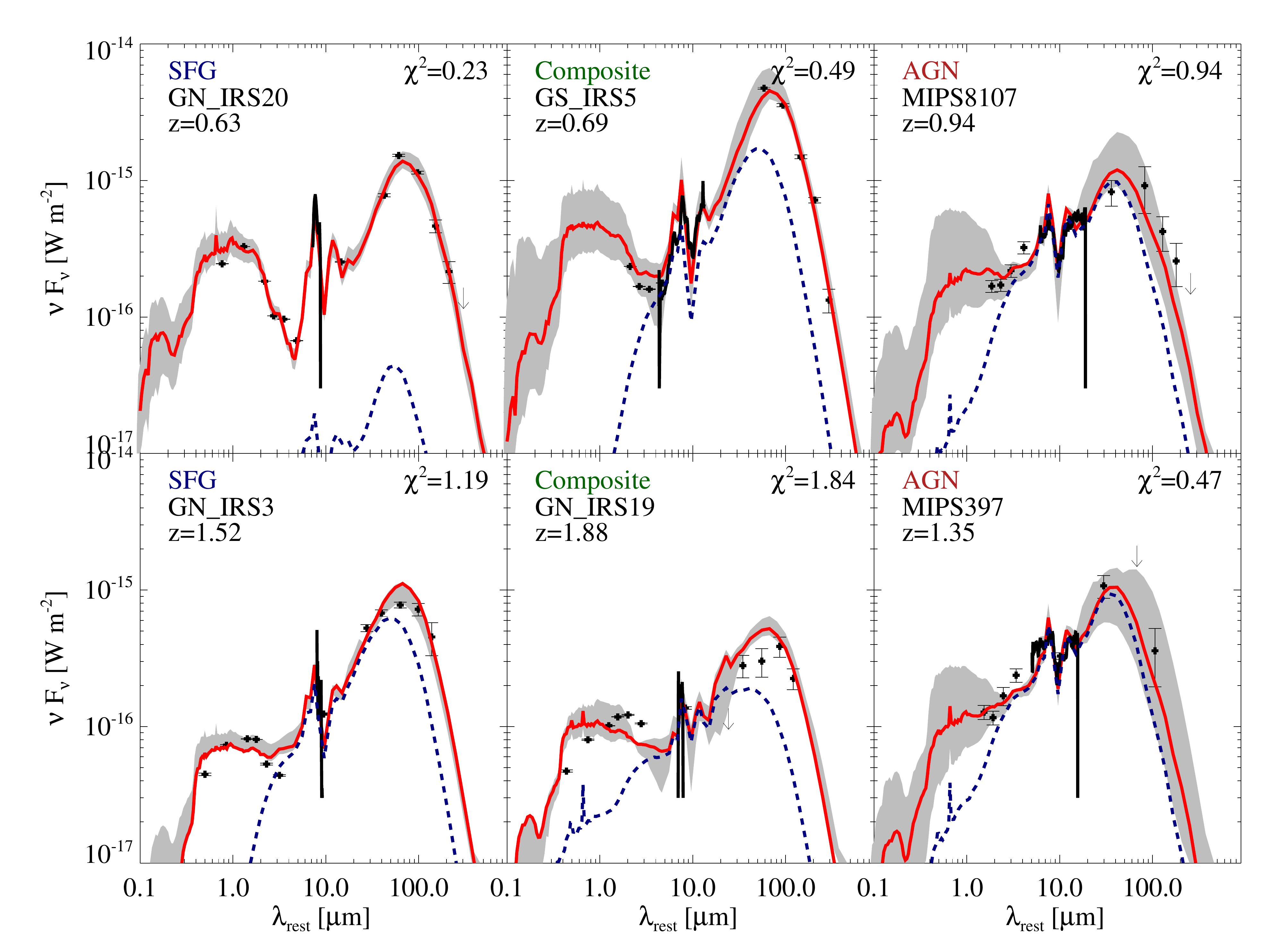} 
   \caption{Example best-fit SEDs for each class in our sample. The black data points are observed photometry, with $2 \sigma$ upper bounds indicated by downward arrows. The error bars represent the observational uncertainties on the photometry. The black solid curves are the observed \emph{Spitzer} IRS spectra. The red solid curves are the best-fit simulated SEDs, and the gray shaded region denotes the spread in SEDs with $\chi^2 < \chi^2_{\rm{min}}+0.5$. The navy dashed curves are the post-processing AGN contribution to the overall SED. The examples presented here, which span a range in AGN fractions, demonstrate that the SEDs forward-modeled from the simulations provide good fits to the observed SEDs.}
   \label{fig:sedmos}
\end{figure*}

\subsection{SED Fitting} \label{sec:fitting} 
The analysis so far has been based on the simulated SEDs alone. We now want to compare our simulated SEDs with the SEDs of our sample of observed galaxies in order to directly compare their simulation-based and empirical SED-fitting based AGN fractions. To find the best-fit simulated SED for each galaxy, we make use of both its broadband photometry and mid-IR spectra. To include the {\sl Spitzer} IRS spectra, we create `pseudo'-photometry using $\Delta \lambda = 2 \ \micron$ square filters \citep{Hernancaballero2007,Sajina2012} at 16 (except where 16 $\rm{\mu m}$ {\sl Spitzer} \emph{peak-up} detections are available{\rcom )}, 18, 20, 22, 26, 28, 30, and 32 $\rm{\mu m}$. This is done to sample the mid-IR without overwhelming the fitting routine or introducing an arbitrary weighting. In addition, this binning of the IRS spectra brings them closer to the spectra resolution of the simulated SEDs themselves (see Figure~\ref{fig:sedmos}). 

For each observed galaxy - model SED pair, we compute a $\chi^2$ value as follows: 
\begin{equation}
\chi^2= \frac{1}{N_{\nu}-1} \sum_{\nu} \frac{(F_{\nu,\data{}}-a\times F_{\nu,\model{}})^2}{\sigma_{\nu,\data{}}^2+\sigma_{\nu, \model{}}^2},
\label{eq:chidef}
\end{equation}
where $N_{\nu}$ is the number of photometric bands, $F_{\nu,\data{}}$ is the observed flux density in each band, $F_{\nu,\rm{model}}$ is the corresponding simulated flux density, $a$ is a linear scale parameter, $\sigma_{\nu,\data{}}$ is the observed uncertainty, and $\sigma_{\nu,\model{}}$ is the estimated model uncertainty, which incorporates the uncertainty associated with fixing certain parameters in the model rather than allowing them to vary (e.g. the dust model); see Paper III (Roebuck et al., in prep.) for details. In cases of missing far-IR data, we additionally constrain our models using the $3 \sigma$ upper limits. When a model exceeds an upper limit for any point past $\lambda_{\rm{obs}}>$\,70\um\, the $\chi^2$ value is multiplied by 100. We only do this for the far-IR points because they determine the overall luminosity of the galaxy; therefore, without this constraint, we can severely overestimate the luminosity and by extension star formation rate or/and obscured AGN luminosity of the galaxy. The model with the lowest $\chi^2$ is taken to be the best-fit model. Including a variable linear scale parameter (see Equation\,\ref{eq:chidef}) achieves overall better $\chi^2$ values. The linear scaling factor $a$ in Equation\,\ref{eq:chidef} is necessary because we have a discrete simulation library that does not fully sample the relevant parameter space and thus likely does not capture the full variation observed in real SEDs. Because use of this scale factor breaks the physical consistency of the model (i.e., the luminosity and shape of the SED are no longer directly physically connected to a specific hydrodynamical simulation), it is best to restrict the scale factor to be only as large as necessary to achieve acceptable fits. A full examination of the effect of this scaling is deferred to Paper III (Roebuck et al., in prep.), but we list the salient conclusions here. We find that when left as a free parameter, the median best-fit scaling factor is 1.73. This suggests that the initial parameters of our simulations library are well suited to this sample. However, the scale factor distribution has broad tails that extend to above or below a factor of 10. We also find that model degeneracies lead to very broad $\chi^2$ vs. model distributions such that a fit with large scale factor may be only marginally better than one with no scale factor (i.e., for which the forward-modeled SED is used). We also re-fit all sources with a limited range of allowed scale factor, $1/3 < a < 3$. We compare the two cases (free vs. restricted scale factor) using the best-fit $\chi^2$ in each case and an odds ratio defined as $P_1/P_2=\exp(-(\chi^2_1-\chi^2_2)/2)$. We find that the odds ratio is typically $>$\,0.95 and at the very worst is 0.67. This means that the two cases are essentially equally likely. Based on the above analysis, in the following, we adopt the restricted-scale-factor fitting.   
Figure~\ref{fig:sedmos} shows representative best-fit SEDs for galaxies from each of the three spectral classes defined in \citet{Kirkpatrick2015}. These SEDs provide visual confirmation of the general goodness of fits obtained for our observed sample using our model library. Overall, the fits are good, with $\sim$94\% achieving $\chi^2<3$. Given that the SEDs are \emph{forward-modeled} from hydrodynamical simulations and a restricted scale factor ($1/3 < a < 3$) is used, achieving fits of this quality is a non-trivial feat. Table\,\ref{tab:bestfitclass} shows the median $\chi^2$ values for each class and their standard deviations. All classes show good fits, although the AGN have marginally worse $\chi^2$ values.

\begin{figure*}
   \includegraphics[width=1.95\columnwidth]{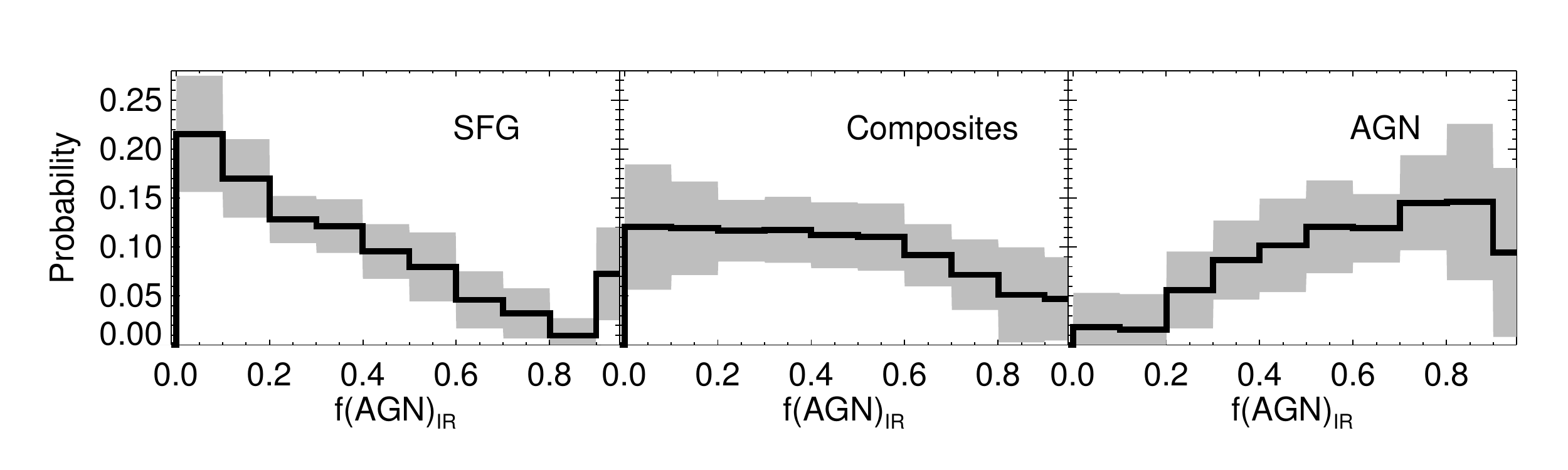} 
   \caption{Probability distributions of \host{} binned by mid-IR classification. Thick black lines are median probabilities, {\rcom grey scale is one median absolute deviation from the median}. These show significant systematic uncertainties on individual estimates, but also show that, as expected, the peak shifts toward larger values with increasing empirical mid-IR AGN strength. 
}
   \label{fig:fagn_mosaic}
\end{figure*}

\begin{figure}
   \centering
   \includegraphics[width=\columnwidth]{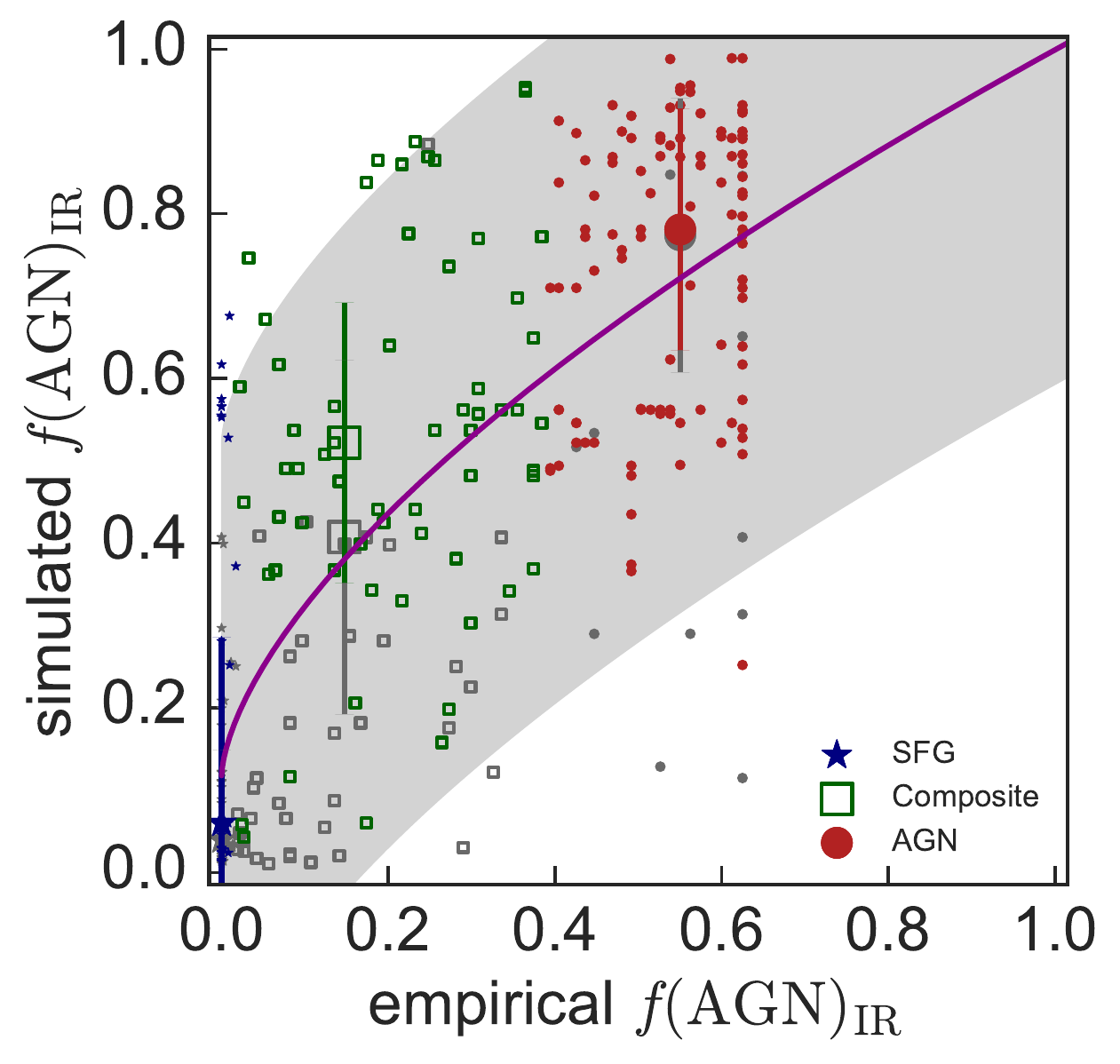} 
   \caption[width=0.95\columnwidth]{Our best-fit \host{} fractions compared with the empirically derived \iremp{} for the different mid-IR spectral classes. The \host{} values for the colored symbols come directly from the simulations, but those for the greyscale symbols use Equation\,\ref{eq:fagnratio} and therefore are more uncertain. The magenta line is the best fit (\host{} = $0.88 \times [$\iremp{}$]^{0.63} + 0.12$, including grey points), {\rcom and the greyscale is 2$\sigma$ to the fit}. The larger symbols and their errorbars represent the median values for each spectral class and median absolute deviation respectively (the greyscale points include all sources within a class, the colored points are restricted to those with direct measurements of \host{}). The broad agreement between the empirically estimated and simulation-based values supports our mid-IR spectral classification, although our simulated fractions imply the AGN contribution to IR-luminous galaxies, especially the Composite galaxies, may be even higher than previously estimated. See the text for discussion on the spread and outliers in this plot.}
   \label{fig:vempiragn}
\end{figure}

\subsection{AGN Fractions From Simulated SED Fits} \label{sec:agnfrac}

Table~\ref{tab:bestfitclass} also shows the median \inp{} and \host{} values as well as the median $A_V$ values of the best-fit simulation SEDs for each of our empirical spectral classes. There is broad agreement with the empirical classification, with SFGs showing the smallest AGN fractions, whereas AGN show the largest AGN fractions. The biggest deviation is seen among the Composite sources where the simulations imply AGN fractions $\sim$\,3\,$\times$ greater than our empirical values. Indeed, our results suggests that for both the Composite and AGN classified sources, the median IR AGN fractions are $>$\,50\%. The median $A_V$ values for all sub-classes place them well within the optically-thick regime, consistent with the similarity between their $\hosteqn$ and $\fagnbol$ values.  

Our simulation-based fits allow us to estimate the systematic uncertainties associated with AGN fraction estimates. This is done by computing a marginalized probability for a particular AGN fraction based on the best $\chi^2$ achieved across all models. In other words, the probability of AGN fraction $i$ scales as $P_i\propto e^{-\chi_{{\rm min},i}^2 /2}$. Figure~\ref{fig:fagn_mosaic} shows these probability distributions for \host{} binned by mid-IR classification. 
{\rcom The median probability distributions for each mid-IR classification are shown as the thick black histograms, with one median absolute deviation given as the greyscale.}
The breadth of these distributions reflects the model degeneracies in that comparable $\chi^2$ values can be achieved with very different AGN fractions. This figure shows that the systematic uncertainties {\rcom are significant, $\sigma_{f{\rm (AGN)_{IR}}} \sim $\,0.4.}

\newpage
\subsection{Comparison with Empirical AGN Fraction Estimates} \label{sec:emp_comparison}

\begin{figure*}
   \includegraphics[scale=0.38]{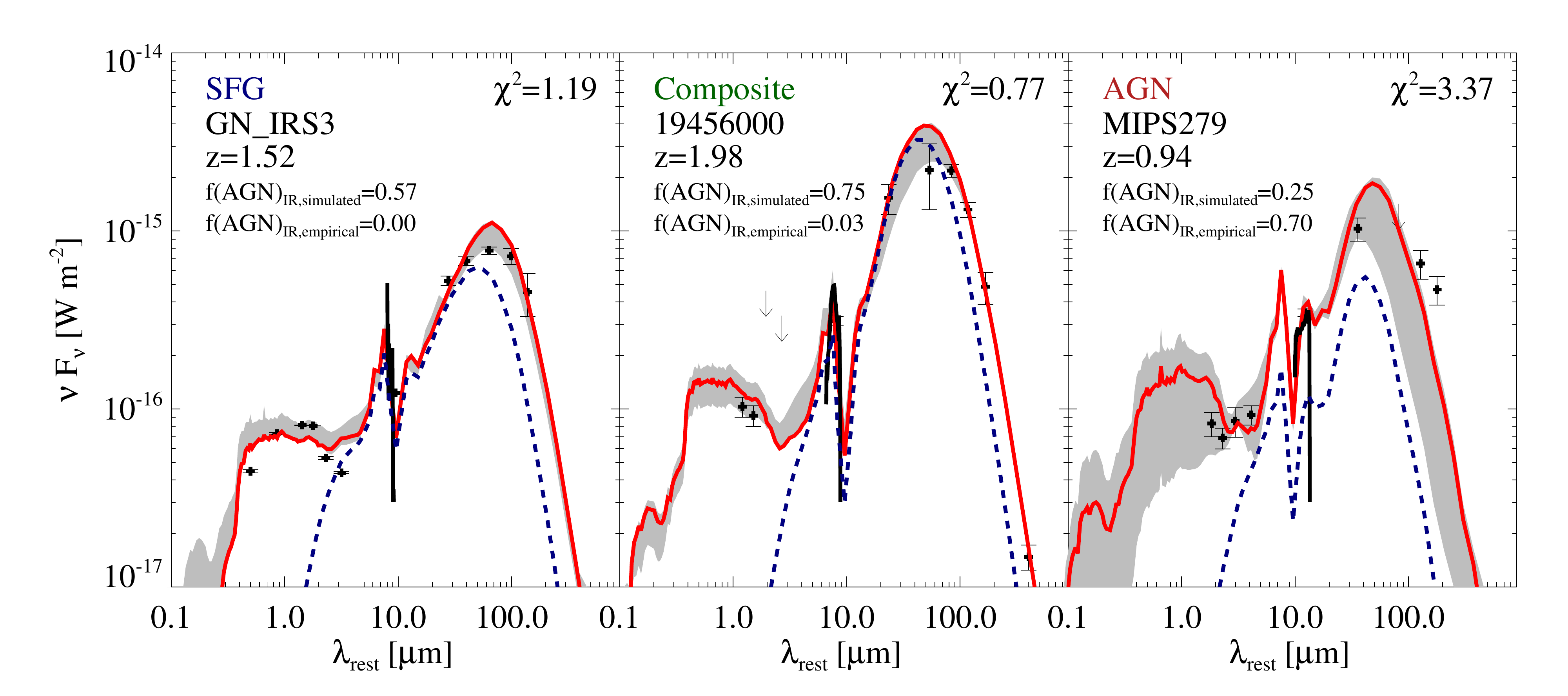} 
   \caption{SEDs of characteristic outliers {\rcom (defined as galaxies at least 2$\sigma$ away from the fit)} from Figure~\ref{fig:vempiragn}. The black data points are observed photometry, with $2 \sigma$ upper bounds indicated by downward arrows. The error bars represent the observational uncertainties on the photometry. The black lines are the observed \emph{Spitzer} IRS spectra. The red curves are the best-fit simulated SEDs, the blue dashed curves are the post host dust processed AGN SEDs, and the gray shaded region denotes the spread in SEDs with $\chi^2 < \chi^2_{\rm{min}}+0.5$ spread. 
 }
   \label{fig:sedmos2}
\end{figure*}

Using the median values, we found broad agreement between our empirical and simulation-based AGN fraction estimates (see Table\,\ref{tab:bestfitclass}). 
Figure~\ref{fig:vempiragn} examines this issue more closely by comparing the \host{} value of the best-fit simulated SED to the empirically derived \iremp{} for each individual galaxy. There is significant scatter, unsurprising given the large systematic uncertainties, but there is a clear trend of larger simulation-based AGN fraction for larger empirical AGN fractions. {\rcom The Spearman rank coefficient is $\rho=$0.75 confirming a strong positive correlation between \host{} and \iremp{}.}

{\rcom The curve shown in Figure\,\ref{fig:vempiragn} is the best-fit power law relation of  
\host{} = $(0.88 \pm 0.10) \times [$\iremp{}$]^{0.63 \pm 0.17} + (0.12 \pm 0.05)$.
The fit was done using {\sc optimize.curve\_fit} from the SciPy library. It was done in linear space assuming uniform $\sigma_{f{\rm (AGN)_{IR}}} \sim $\,0.4 uncertainties estimated from Figure~\ref{fig:fagn_mosaic}. We considered log-space fits as well, however those were complicated by most SFGs having zero \iremp{} values. We caution that, given the large scatter, the above functional form performed only marginally better than a linear fit with a slope close to 1 but a significant offset (i.e. y-intercept) of 0.16. Our preference for the power-law fit is due to both the formally (if marginally) better fit, but also to the better agreement with the median values (the large symbols in Figure\,\ref{fig:vempiragn}).}

{\rcom The above best-fit suggests} that while there is a correlation between the empirical and simulated IR AGN fractions, there is also a significant systematic offset from the expected 1:1 relation. Therefore, our simulations imply the AGN contribution to the IR luminosity of the majority of these sources may be even higher than previously estimated, especially at intermediate empirical AGN fractions (i.e. the composite sources). Nevertheless, the existence of a correlation between the simulation-based and empirical values and the consistency between the empirical and simulation-based classifications (i.e., the sources empirically classified as SFGs have the lowest simulation-based estimates, and those empirically classified as AGN have the highest) suggest that our determinations of the dominant power sources of our IR-selected sources are robust.

Even considering the large scatter, a few points appear to be outliers. {\rcom Here we define as outlier any galaxy whose best-fit $\hosteqn$ value is $\geq$2\,$\sigma$ away from the curve fit seen in Figure~\ref{fig:vempiragn}.} For the SFGs, these are sources where \host{} $\gtrsim$\,0.5, which leads to 6 galaxies (6\% of all SFGs). Nearly all of these sources are fit to simulated galaxies that exhibit strong PAH emission despite having a luminous buried AGN (with $<$\inp{}$> \sim 0.5$). An example of this, GN\_IRS3, can be seen in Figure~\ref{fig:sedmos2}. 

For the AGN, outliers are sources with $\hosteqn$\,$<$0.3, which leads to 5 galaxies (4\% of all AGN). All are $z$\,$<$\,1 galaxies where, for all but one, the IRS spectra does not cover the principle PAH features shortward of the 9.7 \micron{} silicate feature which makes the empirically measured values very uncertain. An example is MIPS279 shown in Figure~\ref{fig:sedmos2}. 

Some of the large scatter in $\hosteqn$ vs \iremp{} for the Composite sources is due to them being subject to both the effects of underestimated deeply obscured AGN (as in the outlier SFGs) as well as the overestimated AGN due to poor IRS coverage (as with the AGN outliers). But Composites also have the broadest median probability distribution (see Figure~\ref{fig:fagn_mosaic}). There are 8 formal outliers (6 above and 2 below the 2\,$\sigma$ band in Figure~\ref{fig:vempiragn}) or $\sim$7\% of all Composites. The ones above are essentially the same as the star-forming galaxy outliers (an example is given in Figure~\ref{fig:sedmos2}; 19456000), whereas the two below are effectively the same as the AGN outliers.

\section{Discussion}  \label{sec:discussion}

\subsection{Caveats Regarding Simulation-Based AGN Fractions}

The spatial resolution of the simulations used here is $\ga 100$ pc, and the ISM tends to be smoother than in reality even on resolved scales because of the use of the \citet{Springel2003} equation of state, which pressurizes the ISM to account for the effects of stellar feedback. The average optical depth through a cell is maximized when the cell is assumed to have constant density; any clumpiness will reduce the mean optical depth, although lines of sight that intersect clumps can have higher optical depths than in the uniform-density case \citep{Witt1996}. We note that $\sim28\%$ of our galaxies are best fit by ``multiphase-on" simulations (see Section~\ref{sec:ism}) - therefore it is possible that the AGN reprocessing is \emph{underestimated} because we simply throw away the dense clumps, which typically account for $\sim 90$\% of the dust mass. However, in Appendix\,B, we examine the role of ``multiphase-on" vs. ``multiphase-off" ISM treatments on the emergent \host{} values and find them to be fully consistent.

This issue was investigated by \citet{Hopkins2012}, who used a multi-scale technique \citep{Hopkins2010} to self-consistently simulate gas inflows from galaxy to AGN `torus' scales. In their Figure 8, \citet{Hopkins2012} demonstrate that assuming that the gas is smooth leads to column densities that are systematically greater than those inferred for real AGN, whereas assuming that the ISM is clumpy on sub-resolution scales leads to lower column densities that are in better agreement with the observationally inferred values. However, we note that \citet{Hopkins2012} used a probabilistic method to account for obscuration from clumps, whereas in our `multiphase on' runs, we simply ignore the clumps. 
We further caution the details of this analysis cannot be translated to our simulations because of significant differences in the resolution, technique, and assumed equation of state for the ISM. Thus, exactly how sub-resolution clumpiness affects the AGN contribution to the IR emission is still uncertain and will be investigated in more detail in the future using simulations with parsec-scale resolution.

That being said, we can empirically judge the degree to which this simulated ``smoothing" of the ISM plays a role here by considering the fact that for SFG and AGN, the simulated and empirical IR AGN fraction estimates agree typically to better than 50\%, although the scatter is significant, partly due to the scatter among camera angles but also model degeneracies. Composite sources show much more discrepant simulated and empirical values (median values differ by 3\,$\times$), but as they are subject to the same ISM treatments as the SFG and AGN, this discrepancy is likely real as opposed to a a by-product of the limitations in our simulations.

Even if we have perfect knowledge of the degree to which an AGN contributes to $L_{\rm{IR}}$, translating this to the overall power balance between AGN and stars is not straightforward. Figure~\ref{fig:agncollage} shows that \host{} relates to \inp{} in a complex manner that depends on both the merger stage and the initial gas-richness of the merger progenitors -- although we remove some of this degeneracy by re-casting these in terms of $A_V$. The \inp{} and \host{} AGN fractions exhibit a $\sim$1:1 relation for the most gas-rich progenitors and only prior to coalescence -- i.e. the highest $A_V$ cases. 

The last caveats relates to the assumed feeding of the black hole and the emergent AGN SED. Our simulations adopt a standard 10\% radiative efficiency and adopt an essentially constant quasar-like AGN SED. Both are appropriate choices for high accretion, radiatively efficient AGN. However, as the gas density drops in the post-coalescence stage, we expect a transition to lower accretion rate, radiatively inefficient AGN \citep[e.g.][]{Best2012}. This does not affect our conclusion that the IR AGN fraction in the high $A_V$ regime is a good proxy for the UV-mm AGN fraction, but it does affect the exact relation between these two fractions in the low obscuration/post-coalescence regime. This will be investigated further in future work (Roebuck et al. in prep.). 

\subsection{Caveats Regarding Empirically Based AGN Fractions}

As we found in Section\,\ref{sec:emp_comparison}, the mid-IR spectral classification works well for $>$\,90\% of both AGN- and SFG-classified sources, and the aforementioned types of outliers constitute a negligible fraction of the sources in these classes. 
The \citet{Kirkpatrick2015} method may underestimate the AGN contribution to the longer wavelength ($\gtrsim 100 \ {\rm \mu m}$) portion of the FIR SED (i.e. it may attribute some AGN-heated-dust emission to star formation). It is very difficult to empirically test this possibility because AGN photons that reach the extended host galaxy are indistinguishable from stellar photons in the FIR.

The caveats regarding the observer-based AGN fractions are effectively highlighted by the outliers in Figure~\ref{fig:vempiragn} and discussed in Section\,\ref{sec:emp_comparison}. Apart from a couple of outliers due to incomplete coverage of the principle PAH features, the outliers suggest that some strong AGN sources may be mistaken for weak ones as a result of the AGN being so heavily obscured that the AGN's mid-IR continuum is reprocessed into the far-IR, leaving behind an SFG-like PAH-dominated mid-IR spectrum. This effect is an issue for 6\,\% of our SFG-classified sources, but is likely behind the much higher median AGN fraction for Composite sources inferred from our analysis compared to the empirical analysis in \citet{Kirkpatrick2015}. Overall, the Composites are the only population whose median \host{} values are significantly discrepant from the \iremp{} values. As discussed above, this is unlikely to be an artifact of the simulations since in that case, we would see it for all classes. 

\citet{Kirkpatrick2015} show that the Composite population represents $\sim$\,30\% of the 24-\um\ source population brighter than 0.1 mJy (comparable to the AGN-classified population). This 24-\um\ depth is fainter than the evolutionary peak in the number counts \citep[e.g.][]{Papovich2004}, and sources above a comparable flux density have been shown to account for the bulk of the cosmic IR background  \citep[CIB;][]{Dole2006}, at its peak. Composite sources are usually not considered when looking at the breakdown of the CIB, and their AGN fraction may be $>$\,50\,\% based on our simulations. This implies that the AGN contribution to the CIB may be underestimated \citep[e.g.][]{Jauzac2011}, {\rcom however the magnitude of this effect is very uncertain at present}. To fully constrain the AGN contribution requires a better understanding of the composite fraction of lower redshift IR sources than are covered in our sample {\rcom and will be addressed in Kirkpatrick et al. in prep.} 

\newpage
\section{Summary \& Conclusions} \label{sec:conclusions}

This paper presents an analysis of the accuracy of and systematic uncertainties inherent in determining the AGN contribution to $L_{\rm{IR}}$ based on fitting IR SEDs as well as the relation between this IR AGN fraction and the bolometric AGN fraction. We used a suite of hydrodynamic simulations on which radiative transfer calculations were performed to yield simulated galaxy SEDs. These simulations were used to investigate the relations between the IR and bolometric AGN fractions and key properties such as merger stage and level of obscuration. The simulated SEDs were then directly fit to the observed IR photometry of a sample of 336 $z$\,$\sim$\,0.3\,--\,2.8, $\log (L_{\rm{IR}}) = 10.4$-$13.7$ galaxies spanning the full range in empirically derived AGN fractions \citep[see][for details]{Kirkpatrick2015}.  Our conclusions are the following:
 
\begin{itemize}
	\item An AGN fraction measured solely in the infrared (here \host{}) is a good predictor of the intrinsic AGN to stellar strength (here \inp{}) but only up to and including coalescence, or conversely while the extinction is high ($A_V \gtrsim 1$). We provide relations to convert empirical IR AGN fraction estimates to bolometric AGN fractions as a function of $A_V$.

	\item Our simulation library well represents our observed sample, as indicated both by the overall goodness of fit (Section~\ref{sec:fitting}) and the examples presented in Figure~\ref{fig:sedmos}. A more extensive discussion will be presented in Paper III (Roebuck et al. in prep.)
	
	\item We provide the first estimate of the systematic uncertainties in deriving the AGN fractions of galaxies. {\rcom We estimate that these uncertainties are significant with typical 1\,$\sigma$ uncertainties of $\sigma_{f{\rm (AGN)_{IR}}} \sim $\,0.4 }.  
		
	\item Within the above uncertainties, there is agreement between our empirically derived and simulation-based IR AGN fractions (i.e. \host{}). Specifically, both the per-class median \host{} values, and the formal fit between individual \host{} and \iremp{} values support our previous classification: i.e. empirically classified SFG have the least AGN contribution to their total power output; empirically classified AGN have the most.

	\item However, in detail there are key differences. For Composite sources, we find a significant shift in that their median empirical IR AGN fraction is $\sim$15\,\%, but we infer $>$\,50\,\% from our simulations. This suggests heavily obscured AGN whose strength is underestimated in empirical methods relying on the observed mid-IR spectra. {\rcom In addition, 6\% of our empirically classified SFGs have AGN fractions $>$ 50\%. Both imply the true number density of luminous AGN may be potentially underestimated.} Given the large systematic uncertainties on our estimates, this result requires independent confirmation.
	
	\item Our empirical AGN fraction estimates rely on an AGN template that is heavy in warm dust emitting at 20-40\um. More common `torus-only' AGN templates that have less emission in this regime, will lead to AGN fraction estimates that are 2$\times$ lower and therefore will lead to much greater disagreement with our simulated AGN fractions. 
	
\end{itemize}

\acknowledgements
We are grateful to the anonymous referee for their careful reading and detailed feedback which improved the content and presentation of this paper. This work is supported by NSF grants AST-1313206 and AST-1312418. C.C.H. is grateful to the Gordon and Betty Moore Foundation for financial support.
This work is based in part on data obtained with the {\sl Spitzer} Space Telescope, which is operated by the Jet Propulsion Laboratory, California Institute of Technology under a contract with NASA. This work also makes use of {\it Herschel} data. {\it Herschel} is an ESA space observatory with science instruments provided by European-led Principal Investigator consortia and with important participation from NASA. 

\bibliography{paperrefs}

\appendix

\section{Role of AGN Template} \label{sec:agntemp}

The key conclusion of this paper is that empirical and simulated IR AGN fraction estimates are broadly consistent. Here we explicitly test to what degree both fractions may or may not be affected by the specific AGN template adopted.

The default AGN SED template in our simulations is that of \citet{Hopkins2007}; hereafter H07. In \citet{Snyder2013}, one of the models in our simulations, M6, was also run with two different choices of AGN template: a face-on ($i=0^{\circ}$) and an edge-on ($i=90^{\circ}$) clumpy torus models from \citet{Nenkova2008}; hereafter N08. In \citet{Snyder2013} we concluded that at high levels of obscuration, the choice of input AGN template does not significantly affect the emergent IR SED. This insensitivity arises because IR-selected sources tend to exhibit high dust columns to the AGN because the sources are selected to be dust-obscured. If a relatively unobscured AGN template, such as H07, is used as input for the radiative transfer calculations, much of the UV--optical light is reprocessed into the IR. This effect will mimic having as input a more obscured AGN template, such as the \citet{Nenkova2008} $i=90^{\circ}$ model, where the IR already accounts for most of the AGN bolometric luminosity. We tested that the emergent \host{} fractions are consistent between the runs with the default (H07) vs. face-on or edge-on \citet{Nenkova2008} models. This is as expected since we already concluded that in the highly obscured regime, \host{} is a good proxy for the intrinsic AGN fraction which depends only on the integrated AGN power and therefore is independent of adopted AGN template. 

\begin{figure}
   \centering
    \includegraphics[width=0.5\columnwidth]{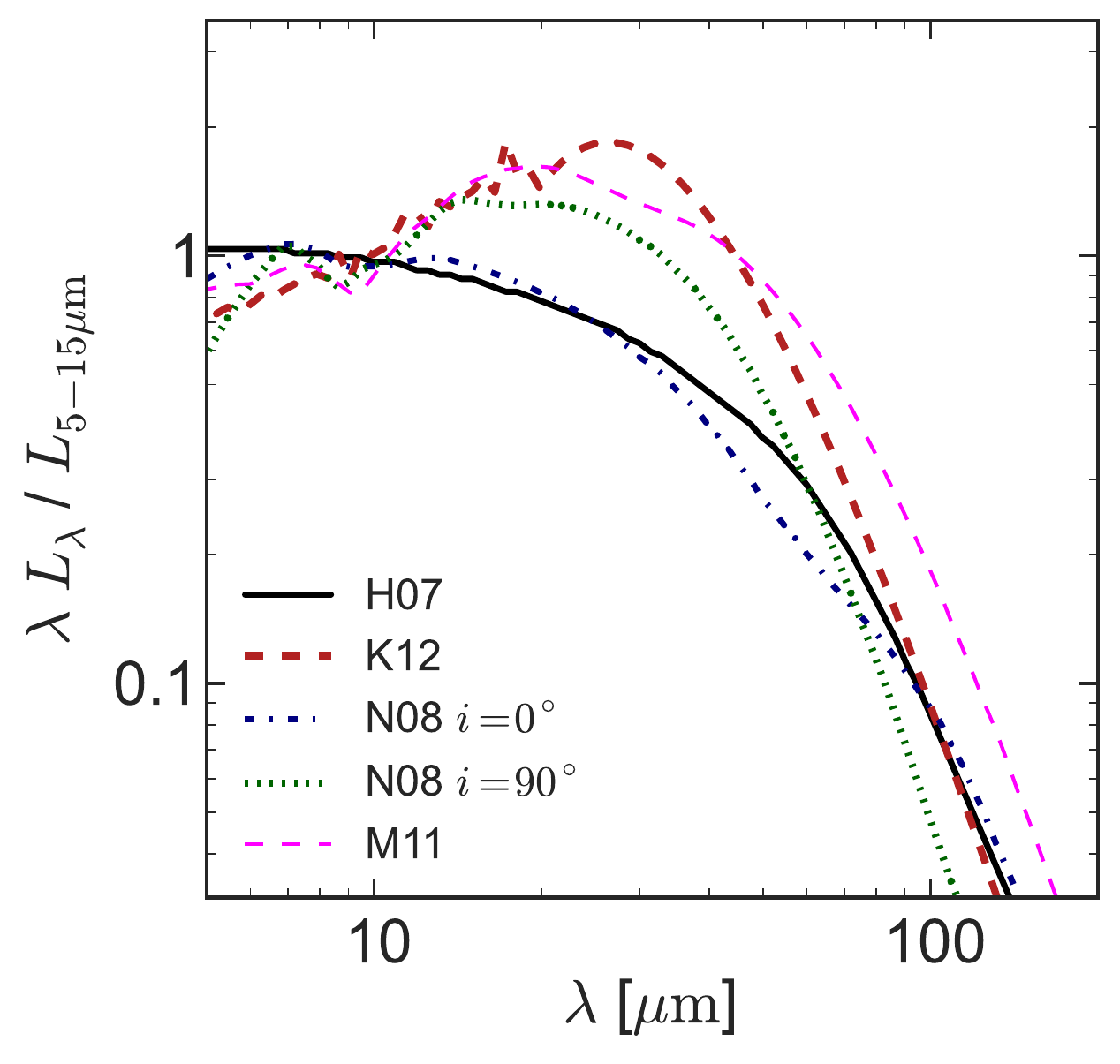} 
      \caption{The range of AGN+torus templates, normalized by their 5-15$\rm{\mu m}$ luminosities. The H07 template (black solid) is the default template in our simulations; while the K12 template (long dashed curve) is the one adopted in our empirical SED decomposition \citep{Kirkpatrick2015}. The templates show significant spread in their level of warm dust ($\sim$\,20\,--\,40\um) emission. In particular, the 8-1000\um\ luminosity ratio between the warm-dust heavy (M11 and K12) templates and warm-dust light templates (face-on N08 and H07) templates is $\sim$\,2. The edge-on N08 templates is intermediate and differs from both the warm-dust heavy and light templates by $\sim$\,50\%. These differences should be born in mind when interpreting other empirical SED decomposition in the literature in the context of the results presented in this paper. See text for full references.} 
         \label{fig:agntemp}
\end{figure}

 By contrast, the empirical IR AGN fraction is very sensitive to the adopted IR AGN template. In \citet{Kirkpatrick2015}, we used the pure AGN template of \citet{Kirkpatrick2012} which is empirically derived from weak silicate absorption AGN with star-formation component subtracted; hereafter the K12 template. A similar method, with consistent results was used in \citet{Mullaney2011} to derive the $z=0$ intrinsic AGN template; hereafter the M11 template\footnote{The phenomenological template adopted in \citet{Sajina2012} was scaled to match M11 template in the long wavelength ``warm-dust" regime, while allowing for the range between face-on and edge-on N08 models in the hot dust regime}. These two templates are compared with the H07 template as well as the edge-on and face-on N08 templates in Figure~\ref{fig:agntemp}. In all cases, we normalize the templates by their 5-15\um\ continuum luminosity to highlight their differences in the ``warm-dust" 20-40\um\ regime. This normalization mimics empirical SED fitting since typically the strength of the AGN relative to star-formation hinges on the level of mid-IR continuum vs. PAH emission -- extrapolating from that to the overall IR AGN fraction is then a function of the AGN template adopted. Clearly, the M11 and K12 are fairly consistent with each other and both show significant warm dust emission. At the other extreme the H07 and face-on N08 models both are much hotter, with weaker 20-40\um\ emission. The difference in integrated 8-1000\um\ AGN luminosity between the top and bottom template in Figure~\ref{fig:agntemp} is 2\,$\times$. Therefore, if we had adopted a face-on N08 model in our empirical SED fitting, our \iremp{} fractions would be lower by 2\,$\times$.  Our conclusion on the agreement between empirical and simulations-based IR AGN fractions is therefore conditional on empirical methods adopting warm-dust heavy templates such as the M11 or K12 ones. 
 
\section{Role of ISM Treatment} \label{sec:agnism}

\begin{figure*}
   \centering
    \includegraphics[width=0.5\columnwidth]{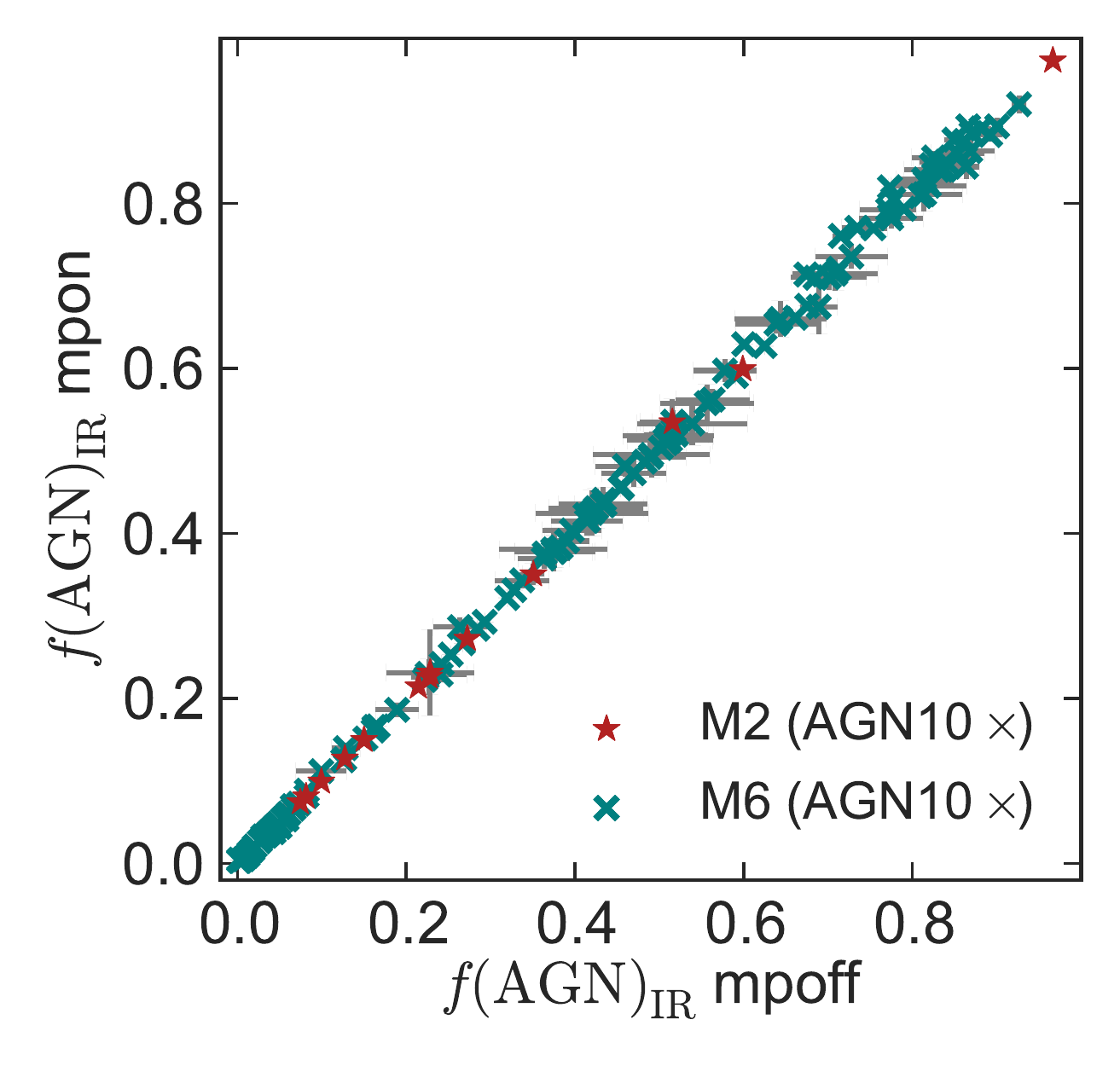} 
   \caption{The \host{} AGN fraction as a function of host ISM treatment. The `multiphase off' ISM treatment (here {\sc mpoff}) assumes that the dust implicit in both phases of the \citet{Springel2003} model is uniformly distributed across each cell and thus provides maximum obscuration. In contrast, the `multiphase on' ISM treatment (here {\sc mpon}) assumes the cold gas has zero volume filling fraction, i.e. the ISM is maximally clumpy (see Sections~\ref{sec:ism}). Teal diamonds represent model \ce{} and red triangles model \lge{} (both AGN10$\times$ mergers). The data points correspond to the means taken over the viewing perspective, and the error bars denote the standard deviations. \host{} does not depend significantly on the ISM assumption because when the IR AGN fractions are high, the AGN are heavily obscured even in the `multiphase on' case.
   }
   \label{fig:mpfagn}
\end{figure*}

Our simulation library is not uniform in terms of the treatment of sub-resolution ISM clumpiness (see Section\,\ref{sec:ism}). Here, we test whether the sub-resolution ISM treatment affects our conclusions. In Figure~\ref{fig:mpfagn}, we plot \host{} corresponding to the fiducial and AGN10$\times$ versions of the M2 and M6 mergers for both the `multiphase on' (i.e. the cold clumps have zero volume filling factor) and ``multiphase off'' (i.e. the ISM is uniform on sub-resolution scales) treatments. These models sample a range of initial gas fractions and AGN strengths. Figure~\ref{fig:mpfagn} suggests that our estimates of \host{} are insensitive to the ISM treatment used. Note that the shape of the IR SED can depend on the sub-resolution ISM assumption because all else being equal, the effective dust temperature of the SED is less in the `multiphase off' case because the dust mass is greater (see \citealt{Lanz2014} and \citealt{Safarzadeh2015} for details). These differences however are lost in our integrated fraction estimates. Ultimately, this result is expected because in the regime in which the IR AGN fraction is non-negligible, the AGN are heavily obscured even in the `multiphase on' case. 

\end{document}